\documentclass[12pt]{article}

\usepackage[dvipsname]{xcolor}

\usepackage{graphicx}
\usepackage{amsmath}        
\usepackage{amssymb}        
\usepackage{slashed}
\usepackage{bm}

\def\mydate{March 11,  2022}
\def\ignore#1{{}}

\def\go{\rightarrow}
\def\dd{\partial}

\def\ep{{\epsilon}}

\def\KK{{\rm KK}}

\def\max{{\rm max}}
\def\onehalf{\hbox{$\frac{1}{2}$}}
\def\la{\langle}
\def\ra{\rangle}
\def\flat{{\rm flat}}
\def\RS{{\rm RS}}
\def\Tr{{\rm Tr} \,}
\def\Psibar{\overline{\Psi}}
\def\diag{{\rm diag}\,}

\def\mybig{\displaystyle \strut }

\def\myfrac#1#2{\frac{\mybig #1}{\mybig #2}}

\def\mymat#1#2{\begin{matrix}#1 \cr \noalign{\kern -2pt} #2\end{matrix}}

\def\mynoalign{\noalign{\kern 4pt}}
\def\mysnoalign{\noalign{\kern 3pt}}

\makeatletter
\@addtoreset{equation}{section}
\makeatother

\begin{document}

\thispagestyle{empty}

{\small \noindent \mydate \hfill OU-HET-1131}

{\small \noindent    \hfill KYUSHU-HET-230}

\vskip 2.cm

\baselineskip=30pt plus 1pt minus 1pt

\begin{center}
{\Large \bf  Anomaly flow by an Aharonov-Bohm phase}

\end{center}


\baselineskip=22pt plus 1pt minus 1pt

\vskip 1.cm

\begin{center}
{\bf Shuichiro Funatsu$^1$, Hisaki Hatanaka$^2$,  Yutaka Hosotani$^3$,}

{\bf Yuta Orikasa$^4$ and Naoki Yamatsu$^5$}

\baselineskip=18pt plus 1pt minus 1pt

\vskip 10pt
{\small \it $^1$Institute for Promotion of Higher Education, Kobe University, Kobe 657-0011, Japan}\\
{\small \it $^2$Osaka, Osaka 536-0014, Japan} \\
{\small \it $^3$Department of Physics, Osaka University, 
Toyonaka, Osaka 560-0043, Japan} \\
{\small \it $^4$Institute of Experimental and Applied Physics, Czech Technical University in Prague,} \\
{\small \it Husova 240/5, 110 00 Prague 1, Czech Republic} \\
{\small \it $^5$Department of Physics, Kyushu University, Fukuoka 819-0395, Japan} \\

\end{center}

\vskip 1.cm
\baselineskip=18pt plus 1pt minus 1pt

\begin{abstract}
In gauge-Higgs unification (GHU), gauge symmetry is dynamically broken by an Aharonov-Bohm (AB) phase,
$\theta_H$, in the fifth dimension.  We analyze $SU(2)$ GHU with an $SU(2)$ doublet fermion 
in flat $M^4 \times (S^1/Z_2)$ spacetime and in the Randall-Sundrum (RS) warped  space.  
With orbifold boundary conditions 
the $U(1)$ part of gauge symmetry remains unbroken at $\theta_H = 0$ and $\pi$.  
The fermion multiplet has chiral zero modes at $\theta_H = 0$, 
which become massive at $\theta_H = \pi$.  In other words chiral fermions are transformed 
to vectorlike fermions  by the AB phase $\theta_H$.  Chiral anomaly 
at $\theta_H = 0$ continuously varies as $\theta_H$ and vanishes at $\theta_H=\pi$.  We 
demonstrate this intriguing phenomenon in the RS space in which there occurs no level crossing 
in the mass spectrum and everything varies smoothly.  
The flat spacetime limit is singular as the AdS curvature of the RS space diminishes, and reproduces 
the result in the flat spacetime.  Anomalies appear for various combinations of Kaluza-Klein 
excitation modes of gauge fields as well. 
Although the magnitude of anomalies depends on $\theta_{H}$ and the warp factor  of the RS space,
it does not depend on the bulk mass parameter of the fermion field controlling its mass and wave function 
at general $\theta_H$.
\end{abstract}


\newpage

\baselineskip=20pt plus 1pt minus 1pt
\parskip=0pt

\section{Introduction} 

In gauge-Higgs unification (GHU), gauge symmetry is dynamically broken by an Aharonov-Bohm (AB) 
phase, $\theta_H$, in the fifth dimension.\cite{Hosotani1983}-\cite{Kubo2002}
In the analysis of finite temperature behavior of grand unified theory (GUT) inspired
$SO(5) \times U(1)_X \times SU(3)_C$ GHU models in the Randall-Sundrum (RS) warped space, 
it has been observed that chiral quarks and leptons 
at $\theta_H = 0$ are transformed to vectorlike fermions at $\theta_H = \pi$.\cite{FiniteT2021} 
As $\theta_H$ varies from 0 to $\pi$, $SU(2)_L \times U(1)_Y \times SU(3)_C$ gauge symmetry is
smoothly converted to $SU(2)_R \times U(1)_{Y'} \times SU(3)_C$ gauge symmetry.
Chiral fermions appearing as zero modes of fermion multiplets in the spinor representation of $SO(5)$
at $\theta_H = 0$ become massive fermions having vectorlike gauge couplings at $\theta_H = \pi$.

Chiral fermions in four dimensions, in general, give rise to chiral anomalies.\cite{Adler1969, BellJackiw1969, Fujikawa1979}
What would be a fate of those anomalies if fermions are converted to massive vectorlike fermions 
at $\theta_H = \pi$?  Do anomalies disappear as $\theta_H$ changes from 0 to $\pi$?
How can it happen?  These are the questions and issues addressed in this paper.

To keep arguments in clarity, 
we analyze $SU(2)$ GHU with an $SU(2)$ doublet fermion both in flat $M^4 \times (S^1/Z_2)$ spacetime 
and in the RS warped  space with orbifold boundary conditions  breaking $SU(2)$ to $U(1)$.  
We shall see that $U(1)$ gauge symmetry survives at $\theta_H = 0$ and $\pi$, 
and that the fermion multiplet has chiral zero modes at $\theta_H = 0$ which become massive at $\theta_H = \pi$.  
We determine 4D  couplings of all Kaluza-Klein (KK) modes of gauge fields and fermion fields at general $\theta_H$,  
and evaluate triangle chiral anomalies.

In the flat $M^4 \times (S^1/Z_2)$ spacetime all gauge couplings are determined analytically, but
the mass spectrum of gauge and fermion fields exhibit level crossing as $\theta_H$ varies.
In the RS spacetime there occurs no level crossing in the spectrum, 
and gauge couplings are evaluated numerically.
It will be seen that 4D gauge couplings of fermions in the RS space smoothly changes as $\theta_H$,
and that the chiral anomaly associated with the zero mode of gauge fields at $\theta_H=0$
smoothly varies and vanishes at $\theta_H = \pi$.
The flat space limit gives singular behavior of the anomaly as a function of $\theta_H$, reproducing
the analytical result in the flat spacetime.
We will also see that anomalies appear in various combinations of KK modes of  gauge fields.

In Section 2 $SU(2)$ GHU models are introduced both in flat  $M^4 \times (S^1/Z_2)$ spacetime
and in the RS space.  As functions of the AB phase $\theta_{H}$ the mass spectra of KK modes
of gauge and fermion fields are obtained.  It will be seen that there occurs no level crossing
in the spectrum in the RS space.  In Section 3 gauge couplings and anomalies are evaluated
in the flat $M^4 \times (S^1/Z_2)$ spacetime.  In Section 4 gauge couplings and anomalies are evaluated
in the RS space.  It is shown that the magnitude of  anomalies smoothly changes as $\theta_{H}$.
The dependence of those anomalies on the warp factor $z_{L}$ of the RS space and the bulk
mass parameter $c$ of fermion fields is also investigated.   It is seen that the flat space limit $z_{L} \go 1$
of anomalies is singular.  It is also seen by numerical evaluation that the magnitude of anomalies 
does not depend on the bulk mass parameter $c$.
Section 5 is devoted to a summary and discussion.

\section{$SU(2)$ GHU models}

The action in $SU(2)$ GHU in flat $M^4 \times (S^1/Z_2)$ spacetime with coordinate 
$x^M$ ($M=0,1,2,3,5$, $x^5 =y$) is given by
\begin{align}
I_\flat &= \int d^4 x \int_0^{L} dy  \, {\cal L}_\flat ~, \cr
\noalign{\kern 5pt}
{\cal L}_\flat & = - \frac{1}{2} \Tr  F_{MN} F^{MN} 
+  \Psibar \gamma^M D_M \Psi +  \Psibar {}' \gamma^M D_M \Psi'  ~,
\label{flataction}
\end{align}
where ${\cal L}_\flat (x^\mu, y) = {\cal L}_\flat (x^\mu, -y) = {\cal L}_\flat (x^\mu, y + 2L)$.
Here $F_{MN} = \dd_M A_N - \dd_N A_N - i g_A [A_M , A_N]$, $A_M = \onehalf  \sum_{a=1}^3 A^a_M \tau^a$
where $\tau^a$'s are Pauli matrices. 
We have introduced two types of $SU(2)$ doublet fermions $\Psi, \Psi'$ with $D_M = \dd_M - i g_A A_M$.
The metric is $\eta_{MN}  = \diag (-1, 1, 1, 1, 1)$ and $\Psibar = i \Psi^\dagger \gamma^0$.
Orbifold boundary conditions are given, with $(y_0, y_1) = (0, L)$, by
\begin{align}
\begin{pmatrix} A_\mu \cr A_y \end{pmatrix} (x, y_j - y) 
&= P_j \begin{pmatrix} A_\mu \cr - A_y \end{pmatrix} (x,  y_j + y) P_j^{-1} ~, \cr
\noalign{\kern 5pt}
\Psi  (x, y_j - y)  &= P_j \gamma^5 \Psi  (x, y_j + y) ~, \cr
\noalign{\kern 5pt}
\Psi'  (x, y_j - y)  &= (-1)^j P_j  \gamma^5  \Psi'  (x, y_j + y) ~, \cr
\noalign{\kern 5pt}
P_0 = P_1 &= \tau^3 ~.
\label{BC1}
\end{align}
The $SU(2)$ symmetry is broken to $U(1)$ by the boundary conditions (\ref{BC1}).
$A^3_\mu, A^{1,2}_y$ are parity even at both $y_0$ and $y_1$, and have constant zero modes.
Let us denote doublet fields as $\Psi = (u, d)^t$ and  $\Psi' = (u', d')^t$.
$u_R$ and $d_L$ are parity even at both $y_0$ and $y_1$, and have zero modes, 
leading to chiral structure.

The KK expansions of gauge fields around the configuration $A_M = 0$ are given, with $L=\pi R$,  by
\begin{align}
A_\mu^{1,2} (x,y) &= \sqrt{\frac{2}{\pi R}} \sum_{n=1}^\infty A_\mu^{1,2 \, (n)} (x) \, \sin \frac{ny}{R} ~, \cr
\noalign{\kern 5pt}
A_\mu^{3} (x,y) &= \frac{1}{\sqrt{\pi R}} A_\mu^{3 \, (0)} (x) +
\sqrt{\frac{2}{\pi R}} \sum_{n=1}^\infty A_\mu^{3 \, (n)} (x) \, \cos \frac{ny}{R} ~, \cr
\noalign{\kern 5pt}
A_y^{1,2} (x,y) &= \frac{1}{\sqrt{\pi R}} A_y^{1,2 \, (0)} (x) +
\sqrt{\frac{2}{\pi R}} \sum_{n=1}^\infty A_y^{1,2 \, (n)} (x) \, \cos \frac{ny}{R} ~, \cr
\noalign{\kern 5pt}
A_y^{3} (x,y) &= \sqrt{\frac{2}{\pi R}} \sum_{n=1}^\infty A_y^{3 \, (n)} (x) \, \sin \frac{ny}{R} ~.
\label{gaugeKKflat1}
\end{align}
Gauge coupling of  4D $U(1)$ gauge fields $A_\mu^{3 \, (0)} (x)$ is given by
\begin{align}
g_4 = \frac{g_A}{\sqrt{L}} ~.
\label{coupling1}
\end{align}
The zero modes $A_y^{1,2 \, (0)}$ may develop a nonvanishing expectation value, which leads to
an AB phase $\theta_H$ along the fifth dimension.
Without loss of generality one may assume that $\la A_y^{1 \, (0)} \ra = 0$.  Then
\begin{align}
&P \exp \bigg\{  i g_A \int_0^{2 L} dy \, \la A_y \ra \bigg\} = e^{ i \theta_H \tau^2 }
= \begin{pmatrix} \cos \theta_H & \sin\theta_H \cr - \sin\theta_H & \cos \theta_H \end{pmatrix}  ~, \cr
\noalign{\kern 5pt}
&\theta_H = g_4 L \, \la  A_y^{2 \, (0)} \ra ~.
\label{ABphase1}
\end{align}
The AB phase $\theta_H$ is a physical quantity.  It couples to fields, affecting their mass spectrum.
It will be shown shortly that the mode expansions in (\ref{gaugeKKflat1}) do not correspond to 
mass eigenstates for $\theta_H \not= 0$ and need to be improved. 

One can change the value of $\theta_H$ by a gauge transformation, which also alters
boundary conditions.  Consider a large gauge transformation given by
\begin{align}
\tilde A_M &= \Omega A_M \Omega^{-1}  + \frac{i}{g_A} \Omega \dd_M \Omega^{-1} ~, ~~
\tilde \Psi = \Omega \Psi ~, ~~ \tilde \Psi'= \Omega \Psi' ~, \cr
\noalign{\kern 5pt}
\Omega &= \exp \Big( \frac{i}{2} \theta (y)  \tau^2 \Big) ~, ~~
\theta (y)  = \theta_H \Big( 1 - \frac{y}{L} \Big)
\label{ABphase2}
\end{align}
under which $\tilde \theta_H = 0$ and boundary condition matrices become
\begin{align}
\tilde P_j &= \Omega( y_j -y) P_j \Omega^{-1} (y_j + y) ~, \cr
\noalign{\kern 5pt}
\tilde P_0 &= \begin{pmatrix} \cos \theta_H & - \sin \theta_H \cr - \sin \theta_H & - \cos \theta_H \end{pmatrix} , ~~
\tilde P_1 = \tau^3 ~.
\label{BC2}
\end{align}
Although the AB phase $\tilde \theta_H$ vanishes, boundary conditions become nontrivial.  
Physics remains the same.  This gauge is called the twisted gauge.\cite{Falkowski2007, HS2007}

In the twisted gauge  $\tilde \theta_H = 0$ so that fields satisfy free equations.  The boundary condition at 
$y =L$ remains the same as in the original gauge so that mode functions take the form 
\begin{align}
\tilde P_{1} = + &: \quad C_{\lambda} (y) = \cos \lambda (y- L) ~, \cr
\tilde P_{1} = - &: \quad  S_{\lambda} (y) = \sin \lambda (y- L) ~.
\label{flatbase1}
\end{align}
At $y=0$, $\tilde A_{\mu}^{1}$ and $\tilde A_{\mu}^{3}$ intertwine with each other.  Their general eigenmodes can be written 
in the form
\begin{align}
\begin{pmatrix} \tilde A_{\mu}^{1} \cr \tilde A_{\mu}^{3} \end{pmatrix} &=
\begin{pmatrix} \alpha S_{\lambda} (y) \cr \beta C_{\lambda} (y) \end{pmatrix}  B_{\mu}^{(\lambda)} (x) ~.
\label{flatgaugewave1}
\end{align}
Note that 
\begin{align}
\begin{pmatrix} A_{\mu}^{1} \cr  A_{\mu}^{3} \end{pmatrix} &=
\begin{pmatrix} \cos \theta (y) & \sin \theta (y) \cr -  \sin \theta (y) &\cos \theta (y) \end{pmatrix}  
\begin{pmatrix} \tilde A_{\mu}^{1} \cr \tilde A_{\mu}^{3} \end{pmatrix} ~.
\label{flatgaugewave2}
\end{align}
Hence the boundary conditions at $y=0$, which may be expressed as $A_{\mu}^{1} (x,0) =0$ and 
$(\dd A_{\mu}^{3} /\dd y)(x,0)=0$,  lead to the condition
\begin{align}
\begin{pmatrix} c_{H} S_{\lambda} (0) & s_{H} C_{\lambda} (0) \cr 
-  s_{H}  S_{\lambda}' (0)  &c_{H} C_{\lambda}' (0) \end{pmatrix}  
\begin{pmatrix} \alpha \cr \beta \end{pmatrix}  = 0 ~, 
\label{flatgaugeBC1}
\end{align}
where $c_{H} = \cos \theta_{H}$ and $s_{H} = \sin \theta_{H}$.
Eigenvalues $\lambda$ must satisfy 
$c_{H}^{2 } S_{\lambda}  C_{\lambda}'  + s_{H}^{2 } C_{\lambda}  S_{\lambda}'  |_{y=0} = 0$, or
\begin{align}
\sin^{2} \lambda \pi R - \sin^{2}\theta_{H} = 0 ~,
\label{flatgaugespectrum1}
\end{align}
which leads to the spectrum for $\lambda$;  $\{  R^{-1} (n \pm \theta_{H}/\pi) \ge 0; \, n : {\rm integers}) \}$.
Zero ($\lambda =0$) modes appear for $\sin \theta_{H}=0$. 
Coefficients $\alpha, \beta$ for each mode are determined by (\ref{flatgaugeBC1}) as well.
KK expansions for $\tilde A_{\mu}^{1}, \, \tilde A_{\mu}^{3}$ are expressed in the form
\begin{align}
\begin{pmatrix} \tilde A_{\mu}^{1} (x,y) \cr \noalign{\kern 5pt} \tilde A_{\mu}^{3} (x,y) \end{pmatrix} 
&= \sum_{n=-\infty}^{\infty } B_{\mu}^{\la n \ra} (x) 
\frac{1}{\sqrt{\pi R}}
\begin{pmatrix} \sin \Big[ \myfrac{ny}{R} - \theta (y) \Big] \cr \cos \Big[ \myfrac{ny}{R} - \theta (y) \Big] \end{pmatrix} .
\label{gaugeKKflat2}
\end{align}
The mass of the $B_{\mu}^{\la n \ra}  (x)$ mode is 
$m_{n} (\theta_{H}) =R^{-1}\big| n + \frac{\theta_{H}}{\pi} \big|$.  
In flat space the KK expansion takes a simpler form in the original gauge;
\begin{align}
\begin{pmatrix} A_{\mu}^{1} (x,y) \cr \noalign{\kern 5pt}  A_{\mu}^{3} (x,y) \end{pmatrix} 
&= \sum_{n=-\infty}^{\infty } B_{\mu}^{\la n \ra}  (x) 
\frac{1}{\sqrt{\pi R}}
\begin{pmatrix} \sin  \myfrac{ny}{R} \cr \cos  \myfrac{ny}{R}\end{pmatrix} .
\label{gaugeKKflat3}
\end{align}
The field $A_{\mu}^{2} (x,y)$  is not affected by $\theta_H$, whose KK expansion  is given by that in (\ref{gaugeKKflat1}).

The fermion field $\Psi$ in the twisted gauge 
\begin{align}
\tilde \Psi = \begin{pmatrix} \tilde u \cr \tilde d \end{pmatrix} 
= \begin{pmatrix} \cos \onehalf \theta(y) & \sin \onehalf \theta(y) \cr
- \sin \onehalf \theta(y) &  \cos \onehalf \theta(y) \end{pmatrix}
\begin{pmatrix} u \cr d \end{pmatrix}
\label{flatfermiwave1}
\end{align}
satisfies  free equations in the bulk region $0 < y < L$ and the original boundary  condition at $y= L$
so that its eigen mode takes the form
\begin{align}
\begin{pmatrix} \tilde u_{R} \cr \tilde d_{R} \end{pmatrix} (x,y) &= 
\begin{pmatrix}  \alpha_{R} C_{\lambda} (y) \cr \beta_{R} S_{\lambda} (y) \end{pmatrix}  f_{\lambda,R} (x) ~, \cr
\noalign{\kern 5pt}
\begin{pmatrix} \tilde u_{L} \cr \tilde d_{L} \end{pmatrix} (x,y) &= 
\begin{pmatrix}  \alpha_{L} S_{\lambda} (y) \cr - \beta_{L} C_{\lambda} (y) \end{pmatrix} f_{\lambda,L} (x) ~, \cr
\noalign{\kern 5pt}
\bar \sigma^{\mu} \dd_{\mu}  f_{\lambda,R} (x) &= \lambda  f_{\lambda,L} (x) ~,~~
\sigma^{\mu} \dd_{\mu}  f_{\lambda,L} (x) = \lambda  f_{\lambda,R} (x) ~,
\label{flatfermiwave2}
\end{align}
where $\sigma^{\mu} = (I_{2}, \vec\sigma )$ and  $\bar\sigma^{\mu} = (-I_{2}, \vec\sigma )$.
It follows from the equations of motion in the bulk that $(\alpha_{R}, \beta_{R}) = (\alpha_{L}, \beta_{L}) $.
The boundary conditions  $(\dd u_{R}/ \dd y)(x,0)=0$ and $d_{R} (x,0) = 0$ lead to
\begin{align}
\begin{pmatrix}  \bar s_{H} C_{\lambda}(0) & \bar c_{H} S_{\lambda} (0) \cr
\bar c_{H} C_{\lambda}' (0) & - \bar s_{H} S_{\lambda}' (0) \end{pmatrix}
\begin{pmatrix} \alpha_{R} \cr \beta_{R} \end{pmatrix} = 0 ~,
\label{flatfermionBC1}
\end{align}
where $\bar c_{H} = \cos \onehalf \theta_{H}$ and $\bar s_{H} = \sin \onehalf \theta_{H}$.
Eigenvalues $\lambda$ must satisfy
\begin{align}
\sin^{2} \lambda \pi R - \sin^{2} \onehalf \theta_{H} = 0 ~,
\label{flatfermionspectrum1}
\end{align}
which leads to the spectrum for $\lambda$; $ \{  R^{-1} (n \pm \theta_{H}/2 \pi) \ge 0; \, n : {\rm integers}) \}$.
Zero ($\lambda =0$) modes appear for $\sin \onehalf \theta_{H}=0$. 
Coefficients $\alpha_{R}, \beta_{R}$ for each mode are determined by (\ref{flatfermionBC1}).
KK expansion for $\tilde \Psi$ is given by
\begin{align}
\begin{pmatrix} \tilde u_{R} (x,y) \cr \noalign{\kern 5pt} \tilde d_{R} (x,y) \end{pmatrix} 
&= \sum_{n=-\infty}^{\infty } \psi_{R}^{\la n \ra}  (x) 
\frac{1}{\sqrt{\pi R}}
\begin{pmatrix} \cos \Big[ \myfrac{ny}{R} - \onehalf \theta (y) \Big] \cr 
 \sin \Big[ \myfrac{ny}{R} - \onehalf \theta (y) \Big] \end{pmatrix}, \cr
\noalign{\kern 5pt}
\begin{pmatrix} \tilde u_{L} (x,y) \cr \noalign{\kern 5pt} \tilde d_{L} (x,y) \end{pmatrix} 
&= \sum_{n=-\infty}^{\infty } \psi_{L}^{\la n \ra}  (x) 
\frac{1}{\sqrt{\pi R}}
\begin{pmatrix} \sin \Big[ \myfrac{ny}{R} - \onehalf \theta (y) \Big] \cr 
- \cos \Big[ \myfrac{ny}{R} - \onehalf \theta (y) \Big] \end{pmatrix}.
\label{fermionKKflat1}
\end{align}
$\psi_{R}^{\la n \ra}  $ and $\psi_{L}^{\la n \ra} $ combine to form
the $\psi^{\la n \ra}  (x) $ mode, whose mass is given by 
$m_{n}(\theta_{H}) = R^{-1} \big| n + \frac{\theta_{H}}{2 \pi} \big|$.  In the original gauge the KK expansion takes
the form
\begin{align}
\begin{pmatrix} u_{R} (x,y) \cr \noalign{\kern 5pt} d_{R} (x,y) \end{pmatrix} 
&= \sum_{n=-\infty}^{\infty } \psi_{R}^{\la n \ra}  (x) 
\frac{1}{\sqrt{\pi R}}
\begin{pmatrix} \cos  \myfrac{ny}{R}  \cr 
 \sin \myfrac{ny}{R}  \end{pmatrix}, \cr
\noalign{\kern 5pt}
\begin{pmatrix}u_{L} (x,y) \cr \noalign{\kern 5pt} d_{L} (x,y) \end{pmatrix} 
&= \sum_{n=-\infty}^{\infty } \psi_{L}^{\la n \ra}  (x) 
\frac{1}{\sqrt{\pi R}}
\begin{pmatrix} \sin \myfrac{ny}{R} \cr 
- \cos  \myfrac{ny}{R}  \end{pmatrix}.
\label{fermionKKflat2}
\end{align}

The KK expansion for the fermion field $\Psi'$ is found in a similar manner.
The spectrum is determined, instead of (\ref{flatfermionspectrum1}), by
\begin{align}
\sin^{2} \lambda \pi R - \cos^{2} \onehalf \theta_{H} = 0 ~,
\label{flatDfermionspectrum1}
\end{align}
leading to the spectrum for $\lambda$; $ \{  R^{-1} (n + \onehalf  \pm \theta_{H}/2 \pi) \ge 0; \, n : {\rm integers}) \}$.
The KK expansion becomes
\begin{align}
\begin{pmatrix} u_{R}' (x,y) \cr \noalign{\kern 5pt} d_{R}' (x,y) \end{pmatrix} 
&= \sum_{n=-\infty}^{\infty } \psi_{R}^{\prime \,{\la n + \frac{1}{2} \ra}  } (x) 
\frac{1}{\sqrt{\pi R}}
\begin{pmatrix} \cos  \myfrac{(n + \onehalf) y}{R}  \cr 
 \sin \myfrac{(n + \onehalf) y}{R}  \end{pmatrix}, \cr
\noalign{\kern 5pt}
\begin{pmatrix}u_{L}' (x,y) \cr \noalign{\kern 5pt} d_{L}' (x,y) \end{pmatrix} 
&= \sum_{n=-\infty}^{\infty } \psi_{L}^{\prime \,  {\la n + \frac{1}{2} \ra} } (x) 
\frac{1}{\sqrt{\pi R}}
\begin{pmatrix} \sin \myfrac{(n + \onehalf) y}{R} \cr 
- \cos  \myfrac{(n + \onehalf) y}{R}  \end{pmatrix}.
\label{fermionKKflat3}
\end{align}
$\ \psi_{R}^{\prime \, {\la n + \frac{1}{2} \ra} }$ and 
$\psi_{L}^{\prime \,{\la n + \frac{1}{2} \ra} }$  combine to form
the $\psi^{\prime \,{\la n + \frac{1}{2} \ra} }(x)$ mode 
with a mass $m_{n + \frac{1}{2}}(\theta_{H}) = R^{-1} \big| n + \onehalf + \frac{\theta_{H}}{2 \pi} \big|$. 

\begin{figure}[tbh]
\centering
\includegraphics[width=100mm]{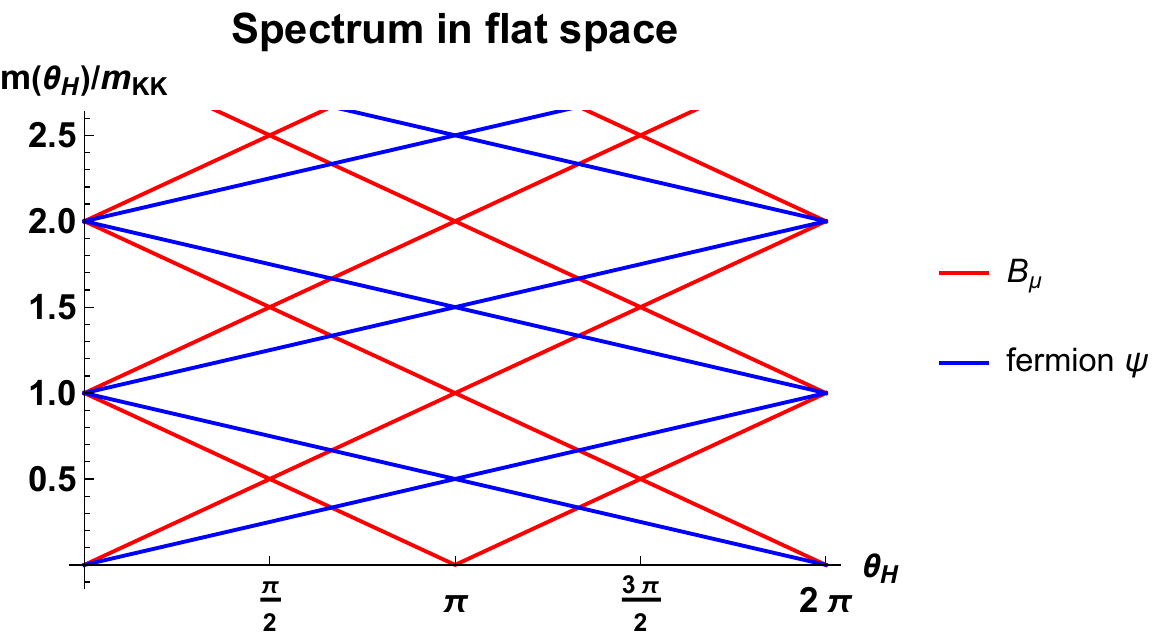}
\caption{The mass spectrum of gauge fields $B_{\mu}^{\la n \ra}$ and fermion fields $\psi^{\la n \ra}$ 
in flat $M^4 \times (S^{1}/Z_{2})$ spacetime is displayed.  Level crossings in the spectrum are seen.
}   
\label{fig:spectrum1}
\end{figure}

The spectrum of $B_{\mu}^{\la n \ra}$ and  $\psi^{\la n \ra}$ 
modes is depicted in Fig.\ \ref{fig:spectrum1} in the range $0 \le \theta_{H} \le 2 \pi$.
The spectrum of $\psi^{\prime \, \la n + \frac{1}{2}\ra}$ modes is obtained
from that of $\psi^{\la n \ra}$ modes by shifting $\theta_H$ by $\pi$.
The KK mass scale is $m_{\KK} = 1/R$ in flat space.
The spectrum of $B_{\mu}$ modes has periodicity with a period $\pi$, whereas the spectrum of
$\psi$ and $\psi'$ modes  has periodicity with a period $2 \pi$.
In flat space the level crossing occurs at $\theta_{H}=0, \onehalf \pi, \pi, \frac{3}{2} \pi$ for $B_{\mu}$,
and  at $\theta_{H}=0, \pi$ for $\psi$ and $\psi'$.

Next we examine $SU(2)$ GHU in the RS space whose metric is given by \cite{RS1999}
\begin{align}
ds^2= e^{-2\sigma(y)} \eta_{\mu\nu}dx^\mu dx^\nu+dy^2
\label{RSmetric1}
\end{align}
where $\eta_{\mu\nu}=\mbox{diag}(-1,+1,+1,+1)$, $\sigma(y)=\sigma(y+ 2L)=\sigma(-y)$ 
and $\sigma(y)=ky$ for $0 \le y \le L$.  It has the same topology as $M^4 \times (S^{1}/Z_{2})$.
In the fundamental region $0 \le y \le L$ the metric can be written, in terms of the conformal coordinate 
$z = e^{ky}$, as
\begin{align}
ds^2=  \frac{1}{z^2} \bigg(\eta_{\mu\nu}dx^{\mu} dx^{\nu} + \frac{dz^2}{k^2}\bigg) 
\quad ( 1 \le z \le z_L= e^{kL}) ~.
\label{RSmetric2}
\end{align}
$z_L$ is called the warp factor of the RS space.
The action in RS is 
\begin{align}
I_\RS &= \int d^5 x  \sqrt{- \det G}   \, {\cal L}_\RS ~, \cr
\noalign{\kern 5pt}
{\cal L}_\RS & = - \frac{1}{2} \Tr  F_{MN} F^{MN} 
+  \Psibar {\cal D} (c) \Psi +  \Psibar {}' {\cal D} (c')  \Psi'  ~, \cr
\noalign{\kern 5pt}
{\cal D} (c) &=  \gamma^A {e_A}^M
\bigg( D_M+\frac{1}{8}\omega_{MBC}[\gamma^B,\gamma^C]  \bigg) - c \, k 
\quad {\rm for} ~ 1 \le z  \le z_{L} ~.
\label{RSaction}
\end{align}
Note ${\cal L}_\RS (x^\mu, y) = {\cal L}_\RS  (x^\mu, -y) = {\cal L}_\RS (x^\mu, y + 2L)$.
Fields $A_{M}$, $\Psi$ and $\Psi'$ satisfy the same boundary conditions (\ref{BC1}) as in the flat space.
The dimensionless bulk mass parameter $c$ in ${\cal D} (c)$ controls the mass and wave function of fermion fields.
The KK mass scale is given by
\begin{align}
m_{\KK} &= \frac{\pi k}{z_{L}-1} 
\label{KKscale1}
\end{align}
which becomes $1/R$ in the flat spacetime limit $k \go 0$.

In the KK expansion $A_{z}^{a} (x,z) = k^{-1/2} \sum A_{z}^{a (n)} (x)  h_{n}(z)$, 
the zero mode  $A_{z}^{2 (0)}$ has a wave function $h_{0} (z) = \sqrt{2/(z_{L}^{2}-1)} \,  z$.
In the $y$-coordinate   $A_{y}^{2 (0)}$ has a wave function
$v_{0} (y) = k e^{ky} h_{0}(z) $ for $0 \le y \le L$ and $v_{0} (-y) = v_{0} (y) =  v_{0} (y+2L)$.
The AB phase $\theta_{H}$ in (\ref{ABphase1}) becomes
\begin{align}
\theta_{H} &= \frac{\la A_{z}^{2 (0)} \ra}{f_{H}} ~, ~~
f_{H} = \frac{1}{g_{4}} \sqrt{ \frac{2k}{L(z_{L}^{2} - 1)} } ~.
\label{ABphase3}
\end{align}
The twisted gauge \cite{Falkowski2007, HS2007}, in which $\tilde \theta_{H} = 0$, is related to the original gauge by a large gauge transformation
\begin{align}
\Omega (z) & = e^{i \theta (z) \tau^{2}/2} ~, ~~
\theta (z) =  \theta_{H} \, \frac{z_{L}^{2} - z^{2}}{z_{L}^{2} - 1} ~.
\label{largeGT1}
\end{align}
In the $y$-coordinate it is written as
\begin{align}
\Omega (y) & = \exp \bigg\{ i \theta_{H} \sqrt{\frac{2}{z_{L}^{2} -1}} \int_{y}^{L} dy \, v_{0} (y) \cdot \frac{\tau^{2}}{2} \bigg\}.
\label{largeGT2}
\end{align}
The boundary conditions in the twisted gauge are given by (\ref{BC2}).

\begin{figure}[tbh]
\centering
\includegraphics[width=100mm]{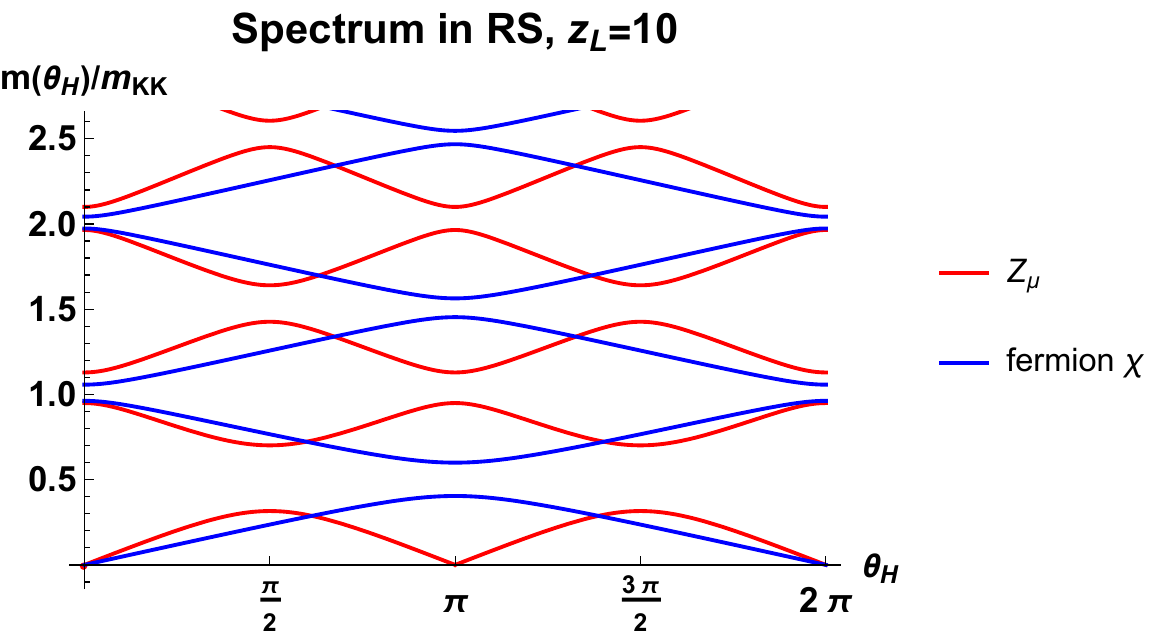}
\caption{The mass spectrum of gauge fields $Z_{\mu}^{(n)}$ and fermion  fields $\chi^{(n)}$
in the RS warped space is displayed.  The warp factor is $z_L = 10$ and the bulk mass parameter 
of $\Psi$ is $c=0.25$.  There is no level crossing in the spectrum.
}   
\label{fig:spectrum2}
\end{figure}

With the boundary conditions at $z=z_{L}$ eigenmodes of $\tilde A_{\mu}^{1}$ and $\tilde A_{\mu}^{3}$ 
are given in the form
\begin{align}
\begin{pmatrix} \tilde A_{\mu}^{1} \cr \tilde A_{\mu}^{3} \end{pmatrix} &=
\begin{pmatrix} \alpha S (z; \lambda) \cr \beta C (z; \lambda) \end{pmatrix}  Z_{\mu}^{(\lambda)} (x) 
\label{RSgaugewave1}
\end{align}
where $S (z; \lambda)$ and  $C (z; \lambda)$ are expressed in terms of Bessel functions and 
are given by (\ref{functionA1}).
The boundary conditions at $z=1$ lead to a condition obtained from (\ref{flatgaugeBC1}) by
substituting $S_{\lambda}(0), C_{\lambda}(0)$ etc.\ by $S (1; \lambda), C (1; \lambda)$ etc.
As $(C S' - SC') (z; \lambda) = \lambda z$, the spectrum is determined by 
\begin{align}
S C' (1; \lambda_{n}) + \lambda_{n} \sin^{2} \theta_{H} = 0 ~.
\label{RSgaugespectrum1}
\end{align}
The corresponding mass is $m_{n} = k \lambda_{n}$.  
We label $\{ \lambda_{n} \}$ from the bottom such that  
$\lambda_{0} (\theta_{H}) < \lambda_{1}  (\theta_{H})< \lambda_{2}  (\theta_{H}) < \cdots$.
There is no level crossing. 
The spectrum is periodic with a period $\pi$, and is displayed in Fig.\ \ref{fig:spectrum2}.

For a fermion field $\Psi(x,z)$ it is most convenient to express its
KK expansion for $\check \Psi(x,z) = z^{-2} \Psi(x,z)$.
Note that Neumann boundary conditions at $z= (z_{0}, z_{1}) = (1, z_{L})$,  
corresponding to even parity,  for left- and right-handed components are
given by
\begin{align}
&D_{+} (c) \check \Psi_{L} \big|_{z_{j}} =0 ~, ~~ D_{-}(c)  \check \Psi_{R} \big|_{z_{j}} =0 ~, ~~ 
D_{\pm}(c)  = \pm \frac{d}{dz} + \frac{c}{z} ~.
\end{align}
In the twisted gauge $\tilde {\check \Psi}$ satisfies  free equations in the bulk.
With the boundary conditions at $z=z_{L}$ eigenmodes of $\check \Psi$ are written in the form
\begin{align}
\begin{pmatrix} \tilde{\check{u}}_{R} \cr \tilde{\check{d}}_{R} \end{pmatrix} (x,z) &= 
\begin{pmatrix}  \alpha_{R} C_{R} (z; \lambda, c) \cr \beta_{R} S_{R} (z; \lambda, c) \end{pmatrix}  f_{\lambda,R} (x) ~, \cr
\noalign{\kern 5pt}
\begin{pmatrix}  \tilde{\check{u}}_{L} \cr\tilde{\check{d}}_{L} \end{pmatrix} (x,z) &= 
\begin{pmatrix}  \alpha_{L} S_{L} (z; \lambda, c) \cr  \beta_{L} C_{L} (z; \lambda, c) \end{pmatrix} f_{\lambda,L} (x) ~, 
\label{RSfermiwave1}
\end{align}
where functions $C_{R/L}$ and $S_{R/L}$ are given in (\ref{functionA3}).
It follows from the equations of motion that $(\alpha_{R}, \beta_{R}) = (\alpha_{L}, \beta_{L})$.
The boundary conditions at $z=1$, $D_{-} \check u_{R}=0$ and $\check d_{R}=0$,  lead to 
\begin{align}
\begin{pmatrix}  \bar s_{H} C_{R}(1) & \bar c_{H} S_{R} (1) \cr
\bar c_{H} S_{L} (1) & - \bar s_{H} C_{L} (1) \end{pmatrix}
\begin{pmatrix} \alpha_{R} \cr \beta_{R} \end{pmatrix} = 0 ~,
\label{RSfermionBC1}
\end{align}
where $C_{R}(1) = C_{R}(1; \lambda, c)$ etc.\ and the relation $D_{-} (c) (C_{R}, S_{R}) = \lambda (S_{L}, C_{L})$
has been used.  As $C_{L} C_{R} - S_{L} S_{R} = 1$, the spectrum is determined by 
\begin{align}
S_{L} S_{R}(1; \lambda_{n}, c) +  \sin^{2} \onehalf \theta_{H} = 0 ~.
\label{RSfermispectrum1}
\end{align}
The corresponding mass is $m_{n} = k \lambda_{n}$.  
As in the case of the gauge field, we label $\{ \lambda_{n} \}$ from the bottom such that 
$\lambda_{0} (\theta_{H}) < \lambda_{1}  (\theta_{H})< \lambda_{2}  (\theta_{H}) < \cdots$.
There is no level crossing.  The  spectrum is periodic with a period $2 \pi$, 
and is displayed in Fig.\ \ref{fig:spectrum2}.

Formulas for a fermion field $\Psi'$ are obtained in a similar manner.  
With the boundary conditions at $z=z_L$ eigenmodes of $\check \Psi'$ are written in the form (\ref{RSfermiwave1}).
The boundary conditions at $z=1$ imply that $\check u_R' =0$ and $D_- \check d_R' = 0$ so that
the spectrum determining equation becomes
\begin{align}
S_{L} S_{R}(1; \lambda_{n}, c') +  \cos^{2} \onehalf \theta_{H} = 0 ~.
\label{RSfermispectrum2}
\end{align}
There appear massless modes at $\theta_H = \pi$.

In Fig.\ \ref{fig:spectrum2} the mass spectrum of gauge fields $Z_\mu^{(n)}$ and fermion fields $\chi^{(n)}$ in RS is depicted.
A distinct feature is that there occurs no level crossing in the RS warped space.
As the AB phase varies, the massless gauge fields $Z_\mu^{(0)}$ at $\theta_H=0$ smoothly changes to 
become massless gauge fields at $\theta_H=\pi$.  The massless mode $\chi^{(0)}$ at $\theta_H=0$, 
on the other hand, becomes massive at $\theta_H=\pi$.  Chiral fermions are transformed to vectorlike
fermions by an AB phase.  We shall confirm it in Section 4 by showing how gauge couplings change as $\theta_H$.
We also stress that the spectrum in the RS space in Fig.\ \ref{fig:spectrum2} converges to the spectrum in $M^4 \times (S^1/Z_2)$ 
in Fig.\ \ref{fig:spectrum1} in the limit $k \go 0$ ($z_L \go 1$).

\section{Anomaly flow in $M^4 \times (S^1/Z_2)$}

4D gauge couplings of fermion fields in flat $M^4 \times (S^1/Z_2)$ spacetime are obtained by
inserting the KK expansions for $A_\mu^{a}$ and $\Psi$ into $-i g_{A} \int d^{4}x dy \, \bar \Psi \gamma^{M} A_{M} \Psi$
and integrating over $y$.
It is most convenient to evaluate the couplings in the original gauge as wave functions of KK modes do not depend on
$\theta_{H}$ in flat spacetime.

The spectrum and KK expansion of  $A_{\mu}^{2} (x,y)$ are not affected by $\theta_{H}$, and therefore
\begin{align}
&- \frac{ig_{A}}{2} \int_{0}^{L} dy \, A_{\mu}^{2} \big\{ u_{R}^{\dagger} \bar \sigma^{\mu} d_{R} - u_{L}^{\dagger} \sigma^{\mu} d_{L}
- d_{R}^{\dagger} \bar \sigma^{\mu} u_{R} + d_{L}^{\dagger} \sigma^{\mu} u_{L} \big\} \cr
\noalign{\kern 5pt}
&= -  \frac{ig_{4}}{2 \sqrt{2}} \sum_{n=1}^{\infty} A_{\mu}^{2 (n)} \sum_{\ell = - \infty}^{\infty} 
\Big\{ \psi_{R}^{\la \ell \ra \, \dagger} \bar \sigma^{\mu} \psi_{R}^{\la \ell +n \ra}
-  \psi_{R}^{\la \ell +n \ra \, \dagger} \bar \sigma^{\mu} \psi_{R}^{\la \ell \ra } \cr
\noalign{\kern 5pt}
&\hskip 4.cm
-\psi_{L}^{\la \ell \ra \, \dagger} \sigma^{\mu} \psi_{L}^{\la \ell +n \ra}
+  \psi_{L}^{\la \ell +n \ra \, \dagger}  \sigma^{\mu} \psi_{L}^{\la \ell \ra}  \Big\} .
\label{flatcoupling1}
\end{align}
Here $\sigma^{\mu}  =( I_{2}, \vec \sigma )$ and  $\bar \sigma^{\mu}  =(- I_{2}, \vec \sigma )$.
All of $ A_{\mu}^{2 (n)}$ couplings  do not depend on $\theta_{H}$ in the above basis.
Note that the zero modes $\psi_{R}^{\la 0 \ra}$ and $\psi_{L}^{\la 0 \ra}$ have
chiral structure in (\ref{fermionKKflat2}).

Couplings of $B_{\mu}^{\la n \ra}$ modes are evaluated similarly.
By inserting the KK expansions (\ref{gaugeKKflat3}) and (\ref{fermionKKflat2}), one finds
\begin{align}
&- \frac{ig_{A}}{2} \int_{0}^{L} dy \, \big\{  A_{\mu}^{1} (\bar u \gamma^{\mu} d +  \bar d \gamma^{\mu} u)
+ A_{\mu}^{3} (\bar u \gamma^{\mu} u -  \bar d \gamma^{\mu} d) \big\}  \cr
\noalign{\kern 5pt}
&= \frac{g_{4}}{2} \sum_{n=-\infty}^{\infty} B_{\mu}^{\la n \ra}  \sum_{\ell =-\infty}^{\infty} 
\Big\{  \psi_{R}^{\la n - \ell \ra\, \dagger} \bar \sigma^{\mu} \psi_{R}^{\la \ell \ra}
+  \psi_{L}^{\la n - \ell \ra \, \dagger}  \sigma^{\mu} \psi_{L}^{\la \ell \ra} \Big\} \cr
\noalign{\kern 5pt}
&=  \frac{g_{4}}{2} \sum_{n=-\infty}^{\infty} B_{\mu}^{\la n \ra}  \sum_{\ell =-\infty}^{\infty} 
\bar{\psi}^{\la n - \ell \ra} \, i \gamma^{5} \gamma^{\mu}  \psi^{\la \ell \ra} ~.
\label{flatcoupling2}
\end{align}
In the $B_{\mu}^{\la n \ra} $ basis the couplings 
$B_{\mu}^{\la n \ra}  \psi_{R}^{\la m \ra \, \dagger}  \psi_{R}^{\la \ell  \ra}$ and 
$B_{\mu}^{\la n \ra}  \psi_{L}^{\la m \ra \, \dagger}  \psi_{L}^{\la \ell \ra}$ take a simple form. 
They are $\onehalf g_{4} \, \delta_{n, m +\ell}$.

At $\theta_{H}=0$ the $n=0$ mode of $B_{\mu}^{\la n \ra}$ is the massless gauge field
of the unbroken $U(1)$.  It has  axial-vector couplings
$\sum_{\ell=-\infty}^{\infty} \bar{\psi}^{\la - \ell \ra}  \, i \gamma^{5} \gamma^{\mu}  \psi^{\la \ell \ra}$.
The coupling to the $\ell =0$ modes leads to a triangle chiral anomaly of three 
$B_{\mu}^{\la 0 \ra}$ legs
with an anomaly coefficient $(g_{4}/2)^{3} (1 + 1)$, reflecting the chiral structure of $u_{R}^{(0)}$ 
and $d_{L}^{(0)}$.  Note that off-diagonal couplings do not contribute to this anomaly.
At $\theta_{H}=\pi$ the   $B_{\mu}^{\la -1 \ra}$ mode becomes the massless gauge field
of the unbroken $U(1)$.  It has  axial-vector couplings
$\sum_{\ell=-\infty}^{\infty} \bar{\psi}^{\la - \ell - 1 \ra}  \, i \gamma^{5} \gamma^{\mu}  \psi^{\la \ell \ra}$.
There arises no chiral anomaly associated with three $B_{\mu}^{\la -1 \ra}$ legs.

To investigate the structure of anomalies let us write $B_{\mu}$ couplings as
\begin{align}
&\frac{g_{4}}{2} \sum_{n=-\infty}^{\infty}  \sum_{m=-\infty}^{\infty}   \sum_{\ell=-\infty}^{\infty} 
B_{\mu}^{\la n \ra}  \Big\{   s^{R}_{n m \ell} \,
\psi_{R}^{\la m \ra \, \dagger} \bar \sigma^{\mu} \psi_{R}^{\la \ell \ra}
+  s^{L}_{n m \ell} \, \psi_{L}^{\la m \ra \, \dagger}  \sigma^{\mu} 
\psi_{L}^{\la \ell \ra} \Big\}  
\label{flatcoupling3}
\end{align}
where, in the current case,  $s^{R}_{n m \ell} = s^{L}_{n m \ell} = \delta_{n, m+\ell}$.
The anomaly coefficient associated with three legs of  
$B_{\mu_{1}}^{\la n_{1} \ra} B_{\mu_{2}}^{\la n_{2} \ra} B_{\mu_{3}}^{\la n_{3} \ra}$ is given by
\begin{align}
&b_{n_{1} n_{2} n_{3}} = b_{n_{1} n_{2} n_{3}}^{R} + b_{n_{1} n_{2} n_{3}}^{L} ~, \cr
\noalign{\kern 5pt}
&b_{n_{1} n_{2} n_{3}}^{R} = \Tr S^{R}_{n_{1}} S^{R}_{n_{2}} S^{R}_{n_{3}} ~,~~
(S^{R}_{n} )_{m \ell} = s^{R}_{n m \ell} ~, \cr
\noalign{\kern 5pt}
&b_{n_{1} n_{2} n_{3}}^{L} = \Tr S^{L}_{n_{1}} S^{L}_{n_{2}} S^{L}_{n_{3}} ~,~~
(S^{L}_{n} )_{m \ell} = s^{L}_{n m \ell} ~.
\label{flatAnomaly1}
\end{align}
It follows that
\begin{align}
&b_{n_{1} n_{2} n_{3}} = \begin{cases} 2 &\hbox{for~}  n_1 + n_2 + n_3 =  \hbox{even}, \cr
\noalign{\kern 5pt}
0 & \hbox{for~}  n_1 + n_2 + n_3 =  \hbox{odd}. \end{cases} 
\label{flatAnomaly2}
\end{align}
Note that  $b_{011}, b_{200}, b_{211}, \cdots =2 \not= 0$.   
Anomalies arise even for massive KK excited gauge bosons in external legs.
Triangle diagrams, for instance,  in which  fermions $\psi^{\la 0 \ra}$,  $\psi^{\la 0 \ra}$ and $\psi^{\la 1 \ra}$ 
($\psi^{\la 1 \ra}$,  $\psi^{\la 1 \ra}$ and $\psi^{\la -1 \ra}$) are running, 
contribute to  $b_{011}$ ($b_{200}$) in perturbation theory.
The divergence of the current $J^{\la n \ra}_\mu$ associated with  $B_{\mu}^{\la n\ra}$ has anomalous terms 
proportional to $\sum_{m \ell} b_{n m \ell} \, \ep^{\mu\nu\rho\sigma} F^{\la m \ra}_{\mu\nu} F^{\la \ell  \ra}_{\rho\sigma} $
where $F^{\la m \ra}_{\mu\nu} = \dd_\mu  B_{\nu}^{\la m \ra}  -\dd_\nu  B_{\mu}^{\la m \ra}$.

In the $B_{\mu}$ basis $b_{n_{1} n_{2} n_{3}} $ is $\theta_{H}$-independent.
The anomaly does not seem to flow with $\theta_{H}$ in the flat space.
However, the level crossing in the spectrum occurs in the flat space.
The $B_{\mu}^{\la 0 \ra}$ mode corresponds to the lowest mode for $0 \le \theta_{H} < \onehalf \pi$,
but becomes the first excited KK mode for $\onehalf \pi < \theta_{H} < \pi$.
In the RS space there is no level crossing.  The lowest gauge field mode remains as the lowest mode for any value
of $\theta_{H}$.  When the AdS curvature of the RS space is very small, namely for $k \ll m_{\KK}$, 
the anomaly associated with three legs of the lowest gauge field must approach to $b_{nnn}$ with $n=-1$
(namely zero) for  $\onehalf \pi < \theta_{H} < \pi$.
In other words the anomaly must flow from 2 to 0 as $\theta_{H}$ varies from 0 to $\pi$.
We are going to see in the next section how this happens.

Contributions of the $\Psi'$ field to anomalies are evaluated in a similar manner.  With the KK expansion 
(\ref{fermionKKflat3}), $B_\mu$ couplings are given by
\begin{align}
\frac{g_{4}}{2} \sum_{n=-\infty}^{\infty}  \sum_{m=-\infty}^{\infty}   \sum_{\ell=-\infty}^{\infty} 
&B_{\mu}^{\la n \ra}  \Big\{   s^{\prime R}_{n m \ell} \,
\psi_{R}^{\prime \la m + \frac{1}{2} \ra \, \dagger} \bar \sigma^{\mu} \psi_{R}^{\prime \la \ell + \frac{1}{2}  \ra}
+  s^{\prime L}_{n m \ell} \, \psi_{L}^{\prime \la m + \frac{1}{2} \ra \, \dagger}  \sigma^{\mu} 
\psi_{L}^{\prime \la \ell + \frac{1}{2} \ra} \Big\}  ~, \cr
\noalign{\kern 5pt}
&{\rm where ~} ~ s^{\prime R}_{n m \ell} = s^{\prime L}_{n m \ell} = \delta_{n, m+\ell +1} ~.
\label{flatcoupling4}
\end{align}
The anomaly coefficient associated with  
$B_{\mu_{1}}^{\la n_{1} \ra} B_{\mu_{2}}^{\la n_{2} \ra} B_{\mu_{3}}^{\la n_{3} \ra}$ is given by
formulae similar to (\ref{flatAnomaly1}) where all quantities are replaced by primed ones, e.g. 
$b_{n m \ell } \go b_{n m \ell }' $ and $s^{R}_{n m \ell} \go s^{\prime R}_{n m \ell}$ etc.
One sees that
\begin{align}
&b_{n_{1} n_{2} n_{3}}' = \begin{cases} 0 &\hbox{for~}  n_1 + n_2 + n_3 =  \hbox{even}, \cr
\noalign{\kern 5pt}
2 & \hbox{for~}  n_1 + n_2 + n_3 =  \hbox{odd}. \end{cases} 
\label{flatAnomaly3}
\end{align}

\section{Anomaly flow in RS}

The KK expansion of gauge fields $A_{\mu}^{1,3}$ becomes
\begin{align}
\begin{pmatrix} \tilde A_{\mu}^{1} (x,z) \cr \noalign{\kern 5pt} \tilde A_{\mu}^{3} (x,z) \end{pmatrix} 
&= \sqrt{k}  \sum_{n=0}^{\infty } Z_{\mu}^{(n)} (x) \, {\bf h}_n (z) ~,
\label{RSgaugeKK1}
\end{align}
where mode functions are given by
\begin{align}
 {\bf h}_0 (z) &= \bar {\bf h}_0^a (z) ~, \cr
\noalign{\kern 5pt}
{\bf h}_{2\ell - 1} (z) &= (-1)^\ell \begin{cases} 
\bar {\bf h}_{2\ell - 1}^a (z) &{\rm for} - \frac{1}{2} \pi  <  \theta_H < \frac{1}{2} \pi \cr
\bar {\bf h}_{2\ell - 1}^b (z) &{\rm for~} 0 < \theta_H < \pi \cr
- \bar {\bf h}_{2\ell - 1}^a (z) &{\rm for~}  \frac{1}{2} \pi < \theta_H < \frac{3}{2} \pi \cr
-\bar {\bf h}_{2\ell - 1}^b (z) &{\rm for~}  \pi < \theta_H < 2 \pi \cr
\bar {\bf h}_{2\ell - 1}^a (z) &{\rm for~} \frac{3}{2} \pi <  \theta_H <  \frac{5}{2} \pi 
\end{cases} ~  ~(\ell = 1, 2, 3, \cdots), 
\cr
\noalign{\kern 5pt}
{\bf h}_{2\ell} (z) &= (-1)^\ell \begin{cases} 
\bar {\bf h}_{2\ell }^b (z) &{\rm for} - \frac{1}{2} \pi  <  \theta_H < \frac{1}{2} \pi \cr
- \bar {\bf h}_{2\ell}^a (z) &{\rm for~} 0 <  \theta_H  < \pi \cr
- \bar {\bf h}_{2\ell}^b (z) &{\rm for~}  \frac{1}{2} \pi < \theta_H < \frac{3}{2} \pi \cr
\bar {\bf h}_{2\ell}^a (z) &{\rm for~}  \pi < \theta_H <  2 \pi \cr
\bar {\bf h}_{2\ell }^b (z) &{\rm for~} \frac{3}{2} \pi <  \theta_H < \frac{5}{2} \pi 
\end{cases}  ~(\ell = 1, 2, 3, \cdots).
\label{RSgaugeKK2}
\end{align}
Here
\begin{align}
&\bar {\bf h}_n^a (z)  
= \frac{1}{\sqrt{r^a_n}} \begin{pmatrix} - s_{H}  \hat S (z; \lambda_n) \cr c_{H} C (z; \lambda_n)   \end{pmatrix} , \cr
\noalign{\kern 5pt}
&\bar {\bf h}_n^b (z)  
= \frac{1}{\sqrt{r^b_n}} \begin{pmatrix} c_{H}   S (z; \lambda_n) \cr s_{H} \check C (z; \lambda_n)   \end{pmatrix} , \cr
\noalign{\kern 5pt}
&r_{n} = \int_{1}^{z_{L}} \frac{dz}{z} \big\{ | h_{n} (z) |^{2} + | k_{n} (z) |^{2} \big\}
\quad {\rm for}~ \begin{pmatrix} h_{n} (z) \cr k_{n} (z) \end{pmatrix} .
\label{RSgaugeKK3}
\end{align}
$\hat S$ and $\check C$ are given in (\ref{functionA2}).
The spectrum determining equation (\ref{RSgaugespectrum1})
can be written as $C S' (1; \lambda_{n}) - \lambda_{n} \cos^{2} \theta_{H} = 0$.  
At $\theta_H=0$ and $\pi$, $S(1; \lambda_{n}) =0$ for even $n$ and  $C'(1; \lambda_{n}) =0$ for odd $n$.
At $\theta_H= \onehalf \pi$ and $ \frac{3}{2} \pi$,  $S' (1; \lambda_{n})=0$ for even $n$ and  $C (1; \lambda_{n})=0$ for odd $n$.
This is why the connection formulas are necessary in (\ref{RSgaugeKK2}).
The expression $\bar {\bf h}_{2\ell -1}^a (z)$, for instance, fails to make sense at $\theta_H = \frac{1}{2} \pi$
as both $\hat S(z; \lambda_{2\ell -1})$ and $c_H$ vanish there.
In deriving the connection formulas, we have made use of an identity 
\begin{align}
&\begin{pmatrix} s_{H}   \hat S (z; \lambda_n) \cr - c_{H} C (z; \lambda_n)   \end{pmatrix}  =
K \begin{pmatrix} c_{H}  S (z; \lambda_n) \cr s_{H}  \check C (z; \lambda_n)   \end{pmatrix} ,  \cr
\noalign{\kern 5pt}
&K =\frac{s_H C(1;\lambda_n)}{c_H S(1;\lambda_n)} =-  \frac{c_H C'(1;\lambda_n)}{s_H S'(1;\lambda_n)} ~ ,
\label{RSidentity1}
\end{align} 
valid at $|\theta_H | \not= 0, \onehalf \pi, \pi , \cdots$.
As a consequence, $\bar {\bf h}_{2\ell - 1}^a (z) =\bar {\bf h}_{2\ell - 1}^b (z)$ for $0 < \theta_H < \onehalf \pi$
and   $\bar {\bf h}_{2\ell - 1}^a (z) = - \bar {\bf h}_{2\ell - 1}^b (z)$ for $\onehalf \pi < \theta_H < \pi$
in (\ref{RSgaugeKK2}).
In numerical evaluation of anomalies we have used, for instance, 
${\bf h}_{2\ell - 1} (z) = (-1)^\ell \bar {\bf h}_{2\ell - 1}^a (z)$ for 
$- \frac{1}{4} \pi < \theta_H \le   \frac{1}{4} \pi$,  $(-1)^\ell \bar {\bf h}_{2\ell - 1}^b (z)$ for 
$ \frac{1}{4} \pi < \theta_H \le   \frac{3}{4} \pi$ and so on.
$ {\bf h}_0 (z)$ is periodic in $\theta_H$ with a period $2\pi$, whereas all other modes $ {\bf h}_n (z)$ ($n \ge 1$)
have a period $\pi$.

Mode functions of the fermion field $\Psi$ are found in a similar manner.
In the KK expansions
\begin{align}
\begin{pmatrix} \tilde{\check u}_{R} (x,z) \cr \noalign{\kern 5pt} \tilde{\check d}_{R} (x,z) \end{pmatrix} 
&= \sqrt{k} \sum_{n= 0}^\infty   \chi_{R}^{(n)} (x) \, {\bf f}_{Rn} (z) ~, \cr
\noalign{\kern 5pt}
\begin{pmatrix} \tilde{\check u}_{L} (x,z) \cr \noalign{\kern 5pt} \tilde{\check d}_{L} (x,z) \end{pmatrix} 
&= \sqrt{k} \sum_{n= 0}^\infty   \chi_{L}^{(n)} (x) \, {\bf f}_{Ln} (z) ~, 
\label{RSfermionKK1}
\end{align}
mode functions are given, for $c>0$,  by
\begin{align}
{\bf f}_{R, 2\ell} (z)  &= 
\begin{cases} \bar {\bf f}_{R, 2\ell}^a (z) &{\rm for~} - \pi < \theta_H < \pi  \cr
\bar {\bf f}_{R, 2\ell}^b (z) &{\rm for~} 0 < \theta_H <  2\pi \cr
- \bar {\bf f}_{R, 2\ell}^a (z) &{\rm for~}   \pi < \theta_H < 3\pi  \cr
- \bar {\bf f}_{R, 2\ell}^b (z) &{\rm for~} 2 \pi < \theta_H <  4\pi \cr
\bar {\bf f}_{R, 2\ell}^a (z) &{\rm for~}  3 \pi < \theta_H < 5 \pi  \cr \end{cases} ~ (\ell=0,1,2, \cdots),\cr
\noalign{\kern 5pt}
{\bf f}_{R, 2\ell-1} (z)  &=  
\begin{cases} \bar {\bf f}_{R, 2\ell-1}^c (z) &{\rm for~} - \pi < \theta_H < \pi  \cr
\bar {\bf f}_{R, 2\ell-1}^d (z) &{\rm for~} 0 < \theta_H <  2\pi \cr
- \bar {\bf f}_{R, 2\ell-1}^c (z) &{\rm for~}   \pi < \theta_H < 3\pi  \cr
- \bar {\bf f}_{R, 2\ell-1}^d (z) &{\rm for~} 2 \pi < \theta_H <  4\pi \cr
\bar {\bf f}_{R, 2\ell-1}^c (z) &{\rm for~}  3 \pi < \theta_H < 5 \pi  \cr \end{cases} ~ (\ell=1,2,3, \cdots) , 
\label{RSfermionKK2R}
\end{align}
and 
\begin{align}
{\bf f}_{L0} (z)  &=  \bar {\bf f}_{L0}^a (z) , \cr
\noalign{\kern 5pt}
{\bf f}_{L, 2\ell-1} (z)  &= 
\begin{cases} \bar {\bf f}_{L, 2\ell-1}^a (z) &{\rm for~} - \pi < \theta_H < \pi  \cr
\bar {\bf f}_{L, 2\ell-1}^b (z) &{\rm for~} 0 < \theta_H <  2\pi \cr
- \bar {\bf f}_{L, 2\ell-1}^a (z) &{\rm for~}   \pi < \theta_H < 3\pi  \cr
- \bar {\bf f}_{L, 2\ell-1}^b (z) &{\rm for~} 2 \pi < \theta_H <  4\pi \cr
\bar {\bf f}_{L, 2\ell-1}^a (z) &{\rm for~}  3 \pi < \theta_H < 5 \pi  \cr \end{cases} ~ (\ell=1,2,3, \cdots) , \cr
\noalign{\kern 5pt}
{\bf f}_{L, 2\ell} (z)  &= 
\begin{cases} \bar {\bf f}_{L, 2\ell}^c (z) &{\rm for~} - \pi < \theta_H < \pi  \cr
\bar {\bf f}_{L, 2\ell}^d (z) &{\rm for~} 0 < \theta_H <  2\pi \cr
- \bar {\bf f}_{L, 2\ell}^c (z) &{\rm for~}   \pi < \theta_H < 3\pi  \cr
- \bar {\bf f}_{L, 2\ell}^d (z) &{\rm for~} 2 \pi < \theta_H <  4\pi \cr
\bar {\bf f}_{L, 2\ell}^c (z) &{\rm for~}  3 \pi < \theta_H < 5 \pi  \cr \end{cases} ~ (\ell=1,2, 3,\cdots),
\label{RSfermionKK2L}
\end{align}
Here
\begin{align}
&\bar {\bf f}_{Rn}^a (z) = \frac{1}{\sqrt{r^a_n}} 
\begin{pmatrix}  \bar c_{H} C_{R}(z; \lambda_{n}, c) \cr - \bar s_{H} \hat S_{R} (z; \lambda_{n}, c)  \end{pmatrix} , ~~
\bar {\bf f}_{Rn}^b (z)  = \frac{1}{\sqrt{r^b_n}} 
\begin{pmatrix}  \bar s_{H} C_{R}(z; \lambda_{n}, c) \cr  \bar c_{H} \check S_{R} (z; \lambda_{n}, c)  \end{pmatrix} ,  \cr
\noalign{\kern 5pt}
&\bar {\bf f}_{Rn}^c (z)  = \frac{1}{\sqrt{r^c_n}} 
\begin{pmatrix}  \bar s_{H} \hat C_{R}(z; \lambda_{n}, c) \cr  \bar c_{H}  S_{R} (z; \lambda_{n}, c)  \end{pmatrix} , ~~
\bar {\bf f}_{Rn}^d (z)  = \frac{1}{\sqrt{r^d_n}} 
\begin{pmatrix}  - \bar c_{H} \check C_{R}(z; \lambda_{n}, c) \cr  \bar s_{H}  S_{R} (z; \lambda_{n}, c)  \end{pmatrix} ,\cr
\noalign{\kern 5pt}
&\bar {\bf f}_{Ln}^a (z)  = \frac{1}{\sqrt{r^a_n}} 
\begin{pmatrix} \bar s_{H} \hat S_{L}(z; \lambda_{n}, c) \cr \bar c_{H} C_{L} (z; \lambda_{n}, c) \end{pmatrix} , ~~
\bar {\bf f}_{Ln}^b (z)  = \frac{1}{\sqrt{r^b_n}} 
\begin{pmatrix} - \bar c_{H} \check S_{L}(z; \lambda_{n}, c) \cr \bar s_{H} C_{L} (z; \lambda_{n}, c) \end{pmatrix} , \cr
\noalign{\kern 5pt}
&\bar {\bf f}_{Ln}^c (z)  = \frac{1}{\sqrt{r^c_n}} 
\begin{pmatrix} \bar c_{H} S_{L}(z; \lambda_{n}, c) \cr  - \bar s_{H} \hat C_{L} (z; \lambda_{n}, c) \end{pmatrix} , ~~
\bar {\bf f}_{Ln}^d (z)  = \frac{1}{\sqrt{r^d_n}} 
\begin{pmatrix}  \bar s_{H}  S_{L}(z; \lambda_{n}, c) \cr \bar c_{H} \check C_{L} (z; \lambda_{n}, c) \end{pmatrix} , \cr
\noalign{\kern 5pt}
&\hbox{where} ~  r_n = \int_1^{z_L} dz  \big\{ | f_{n} (z) |^{2} + | g_{n} (z) |^{2} \big\} 
\quad {\rm for}~ \begin{pmatrix} f_{n} (z) \cr g_{n} (z) \end{pmatrix} .
\label{RSfermionKK3}
\end{align}
Functions $\hat S_{R/L}, \check S_{R/L}$ etc. are defined in (\ref{functionA4}).
At $\theta_H = 0$ the $n=0$ mode is massless; $\lambda_0 =0$.  Its wave function has chiral structure.
$\chi_R^{(0)}$ is $u$-type, whereas $\chi_L^{(0)}$ is $d$-type.  It is seen below that the $n=0$ mode
becomes vectorlike as $\theta_H$ varies to $\pi$.
At $\theta_H=0$, $S_L(1; \lambda_{n}, c) =0$ for even $n$ whereas $S_R(1; \lambda_{n}, c) =0$ for odd $n$.
At $\theta_H=\pi$, $C_R(1; \lambda_{n}, c) =0$ for even $n$ whereas $C_L(1; \lambda_{n}, c) =0$ for odd $n$.
This is why the connection formulas are necessary in (\ref{RSfermionKK2R}) and (\ref{RSfermionKK2L}).
The wave function of the $\chi_{L}^{(0)}$ mode,  ${\bf f}_{L0} (z)$,  is periodic in $\theta_H$ with a period $4\pi$.
Wave functions of all other modes are periodic in $\theta_H$ with a period $2\pi$.
Wave functions for $c <0$ are tabulated in Appendix B.

The four-dimensional part of the gauge interactions for the $\Psi $ field is given by
\begin{align}
g_A \int d^4 x \int_1^{z_L} \frac{dz}{k} \, \Big\{ 
\tilde{\check \Psi}_R^\dagger \bar \sigma^\mu \tilde A_\mu \tilde{\check \Psi}_R
- \tilde{\check \Psi}_L^\dagger \sigma^\mu \tilde A_\mu \tilde{\check \Psi}_L \Big\} ~.
\label{RSgagueInt1}
\end{align}
To find the $Z_\mu^{(n)}$ couplings of fermion modes, we write
\begin{align}
&{\bf h}_n (z) = \begin{pmatrix} h_n(z) \cr k_n(z) \end{pmatrix} , ~~
{\bf f}_{Rn} (z) = \begin{pmatrix} f_{Rn}(z) \cr g_{Rn}(z) \end{pmatrix} , ~~
{\bf f}_{Ln} (z) = \begin{pmatrix} f_{Ln}(z) \cr g_{Ln}(z) \end{pmatrix} .
\label{RSwave1}
\end{align}
By inserting  the KK expansions (\ref{RSgaugeKK1}) and (\ref{RSfermionKK1}) into (\ref{RSgagueInt1}), 
the $Z_\mu^{(n)}$ couplings of the $\chi_n$ fields are found to be
\begin{align}
&\frac{g_4}{2} \sum_{n=0}^\infty \sum_{\ell = 0}^\infty \sum_{m=0}^\infty Z_\mu^{(n)} (x)
\Big\{ t^R_{n\ell m} \, \chi_R^{(\ell)} (x)^\dagger \bar \sigma^\mu \chi_R^{(m)} (x) 
+ t^L_{n\ell m} \, \chi_L^{(\ell)} (x)^\dagger  \sigma^\mu \chi_L^{(m)} (x) \Big\} , \cr
\noalign{\kern 5pt}
&\hskip 1.cm 
t^R_{n\ell m} = \sqrt{kL} \int_1^{z_L} dz \, \Big\{ 
h_n(z) \big( f_{R\ell}^* (z) g_{R m} (z)  + g_{R\ell}^* (z) f_{R m} (z) \big) \cr
\noalign{\kern 5pt}
&\hskip 4.5cm
+ k_n(z) \big( f_{R\ell}^* (z) f_{R m} (z)  - g_{R\ell}^* (z) g_{R m} (z) \big) \Big\} , \cr
\noalign{\kern 5pt}
&\hskip 1.cm
t^L_{n\ell m} =  - \sqrt{kL} \int_1^{z_L} dz \, \Big\{ 
h_n(z) \big( f_{L\ell}^* (z) g_{L m} (z)  + g_{L\ell}^* (z) f_{L m} (z) \big) \cr
\noalign{\kern 5pt}
&\hskip 4.5cm
+ k_n(z) \big( f_{L\ell}^* (z) f_{L m} (z)  - g_{L\ell}^* (z) g_{L m} (z) \big) \Big\} .
\label{RSgaugeCoupling1}
\end{align}
The anomaly coefficient associated with three legs  of $Z_{\mu_1}^{(n_1)} Z_{\mu_2}^{(n_2)} Z_{\mu_3}^{(n_3)}$
is given by 
\begin{align}
&a_{n_{1} n_{2} n_{3}} = a_{n_{1} n_{2} n_{3}}^{R} + a_{n_{1} n_{2} n_{3}}^{L} ~, \cr
\noalign{\kern 5pt}
&a_{n_{1} n_{2} n_{3}}^{R} = \Tr T^{R}_{n_{1}} T^{R}_{n_{2}} T^{R}_{n_{3}} ~,~~
(T^{R}_{n} )_{m \ell} = t^{R}_{n m \ell} ~, \cr
\noalign{\kern 5pt}
&a_{n_{1} n_{2} n_{3}}^{L} = \Tr T^{L}_{n_{1}} T^{L}_{n_{2}} T^{L}_{n_{3}} ~,~~
(T^{L}_{n} )_{m \ell} = t^{L}_{n m \ell} ~.
\label{RSAnomaly1}
\end{align}
Unlike $s^{R/L}_{n m \ell}$ in the flat space, $t^{R/L}_{n m \ell} $ is $\theta_{H}$-dependent.
The anomaly coefficient $a_{n_{1} n_{2} n_{3}}$ also becomes $\theta_{H}$-dependent, exhibiting 
the anomaly flow.

Let us first examine $a_{000} (\theta_{H})$ at $\theta_{H}=0$ and $\pi$, where the gauge field $Z_{\mu}^{(0)}$ 
becomes massless.  At $\theta_{H}=0$, $h_{0}(z) = 0$ and $k_{0} (z) = 1/\sqrt{kL}$.  The fermion zero mode is chiral,
$g_{L0},  f_{R0}  \not= 0$ and $g_{R0} = f_{L0} =0$.  
All other modes are vectorlike; $f_{R\, 2\ell -1} = f_{L\, 2\ell -1} = g_{R\, 2\ell} = g_{L\, 2\ell} = 0$
for $\ell = 1, 2, 3 , \cdots$.   It follows from the ortho-normality conditions that 
$t^{R}_{0 m \ell} , t^{L}_{0 m \ell} =0$ for $m \not=  \ell$.
It is seen that $t^{R}_{0 00} =  t^{L}_{0 00} =1$ and $t^{R}_{0 nn} = - t^{L}_{0 nn} = (-1)^{n}$ for $n \ge 1$.
Hence $a_{000} (0) = 2$, which is the same value as $b_{000} (0) =2$ in the flat space.
 
At $\theta_{H}=\pi$, $h_{0}(z) = 0$ and $k_{0} (z) = - 1/\sqrt{kL}$.  All of the fermion modes are vectorlike;
$g_{R\, 2\ell} = g_{L\, 2\ell} = f_{R\, 2\ell +1} = f_{L\, 2\ell +1} = 0$
for $\ell = 0, 1, 2, \cdots$. Further $t^{R}_{0 m \ell} , t^{L}_{0 m \ell} =0$ for $m \not=  \ell$, and
$- t^{R}_{0 nn} = t^{L}_{0 nn} = (-1)^{n}$ for $n \ge 0$.  It follows that  $a_{000} (\pi) = 0$,
which agrees with $b_{-1, -1, -1} (\pi) =0$ in the flat space.

\begin{figure}[tbh]
\centering
\includegraphics[height=60mm]{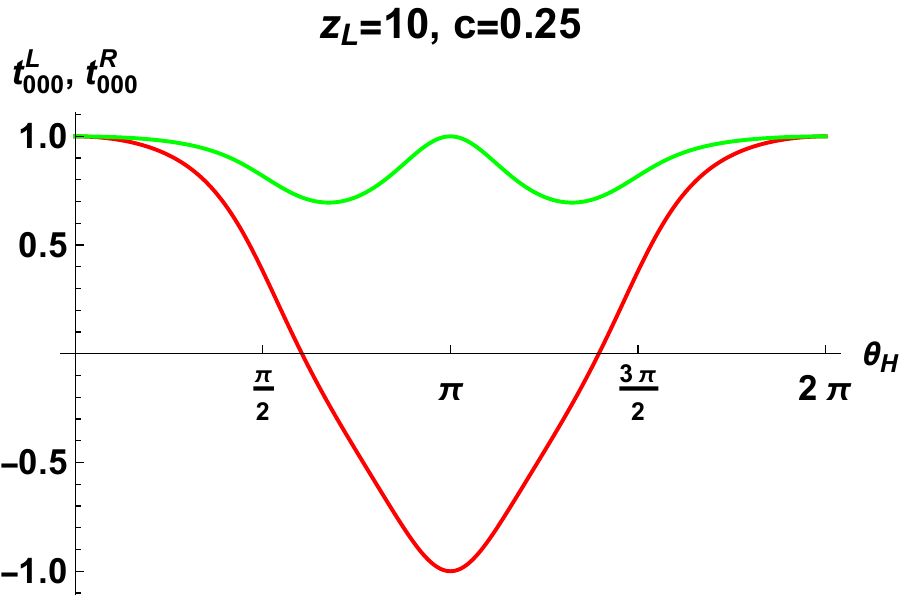}
\caption{The coupling constants $t^R_{000} (\theta_H)$ (red) and $t^L_{000}  (\theta_H)$ (green) 
in (\ref{RSgaugeCoupling1})  are displayed  for the warp factor $z_L = 10$  and the bulk mass parameter  $c=0.25$.
}   
\label{fig:coupling1}
\end{figure}

\begin{figure}[tbh]
\centering
\includegraphics[height=60mm]{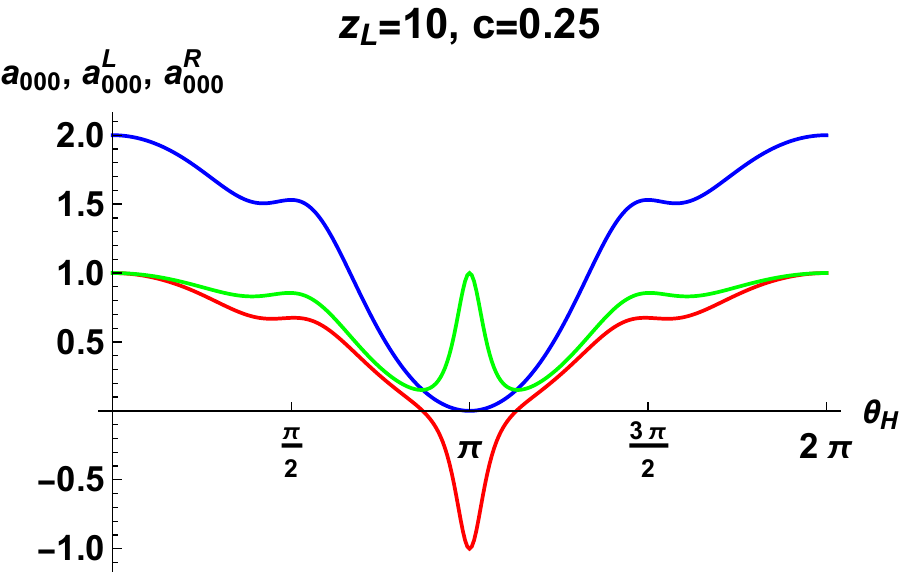}
\caption{The anomaly coefficients $a_{000} (\theta_H)$ (blue), $a^R_{000} (\theta_H)$ (red) and $a^L_{000}  (\theta_H)$ 
(green) in (\ref{RSAnomaly1}) are displayed 
 for the warp factor $z_L = 10$  and the bulk mass parameter  $c=0.25$.
}   
\label{fig:anomaly1}
\end{figure}

The $\theta_H$-dependence of the coupling constants $t^R_{000}$, $t^L_{000} $ and 
anomaly coefficients $a_{000}$, $a^R_{000}$, $a^L_{000}$ is displayed in Figs.\ \ref{fig:coupling1}
and \ref{fig:anomaly1} for $z_L=10$ and $c=0.25$. 
All of them smoothly changes as $\theta_H$.  
The coupling constants of the fermion zero modes are maximally chiral at $\theta_H=0$, 
but become purely vectorlike at $\theta_H=\pi$.
The anomaly is exactly cancelled among the right-handed and left-handed components at $\theta_H=\pi$.
We note that for the anomaly coefficient $a_{n_{1} n_{2} n_{3}} $ off-diagonal gauge couplings
$t^{R/L}_{n m \ell}$ also contribute in (\ref{RSAnomaly1}).
In the previous section we have seen that in the flat space  off-diagonal gauge couplings
$s^{R/L}_{n m \ell} = \delta_{n, m + \ell}$ are important to $b_{n_{1} n_{2} n_{3}} $.  
In the RS space the couplings $t^{R/L}_{n m \ell} (\theta_H)$ are more involved, giving rise to the nontrivial 
$\theta_H$-dependence of $a_{n m \ell}$.

\begin{figure}[bth]
\centering
\includegraphics[width=70mm]{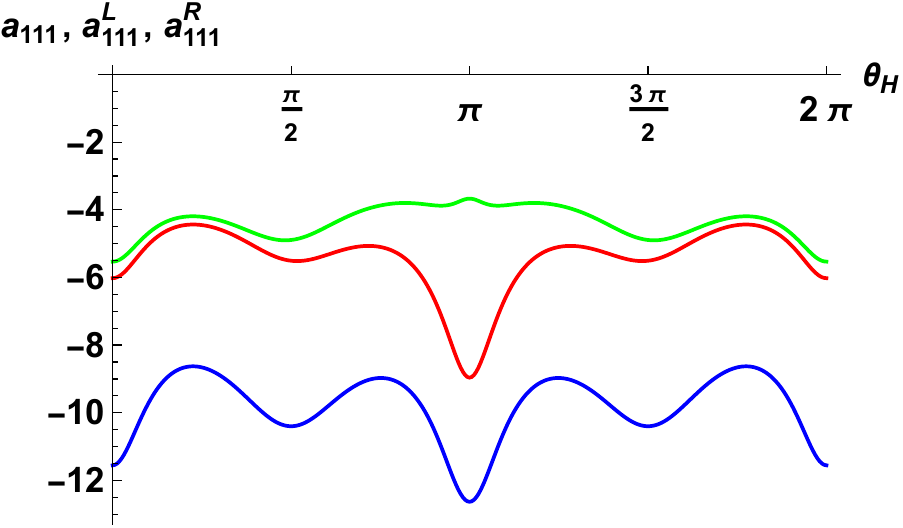}
~
\includegraphics[width=70mm]{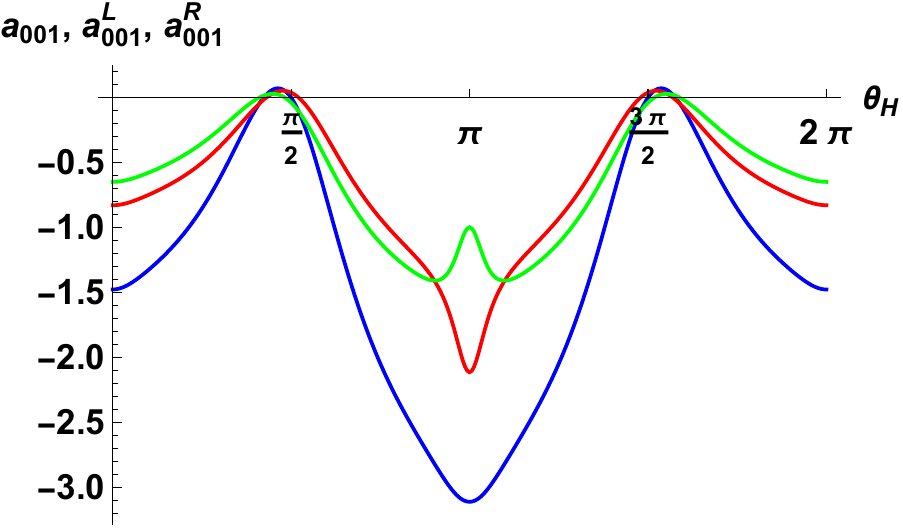}
\vskip 8pt
\includegraphics[width=70mm]{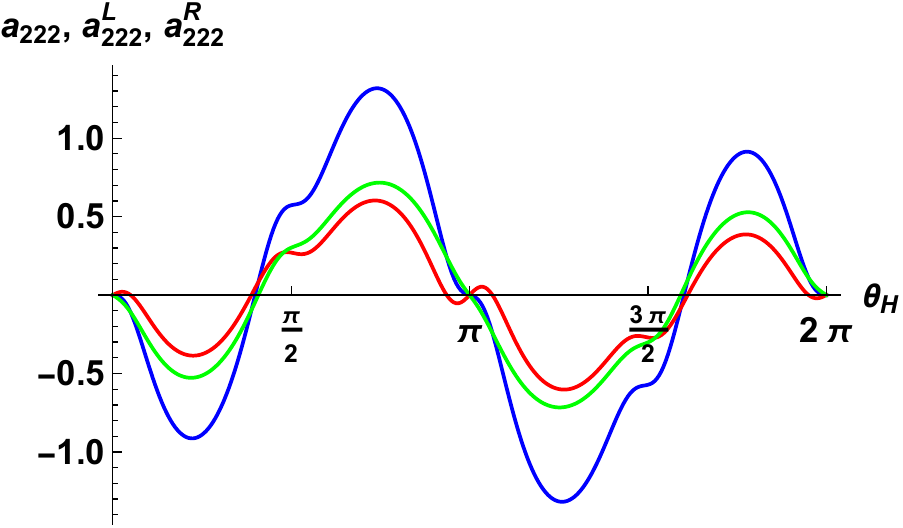}
~
\includegraphics[width=70mm]{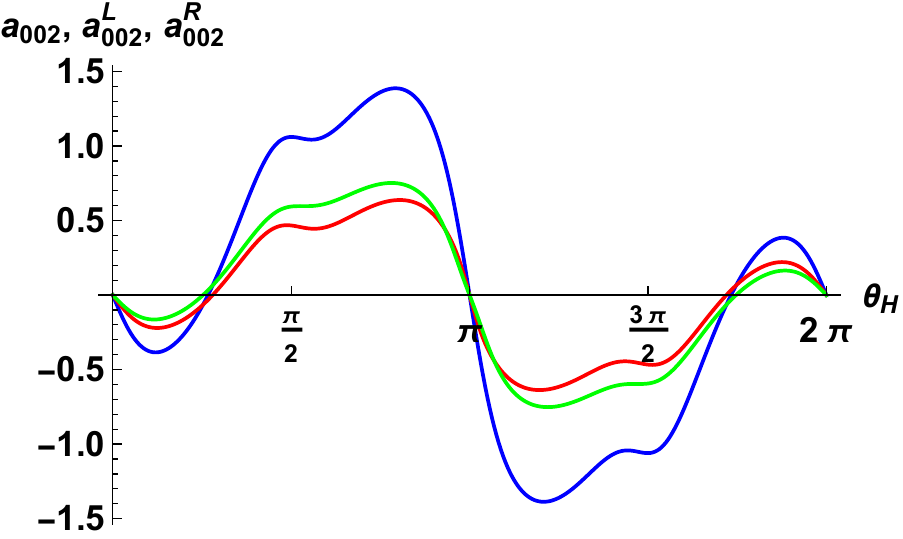}
\caption{The anomaly coefficients $a_{111} (\theta_H)$, $a_{001} (\theta_H)$, $a_{222} (\theta_H)$,
$a_{002} (\theta_H)$  in (\ref{RSAnomaly1})  are displayed  for the warp factor $z_L = 10$  and
the bulk mass parameter  $c=0.25$.
Blue, red, and green lines correspond to $a_{n \ell m}$, $a_{n \ell m}^R$, and $a_{n \ell m}^L$, respectively.
}   
\label{fig:anomaly2}
\end{figure}

Anomalies appear for various combinations of external gauge fields.
In Fig.\ \ref{fig:anomaly2} anomalies $a_{111} $, $a_{222} $, $a_{001} $, and $a_{002} $
are displayed.
In the RS space gauge couplings of the first excited gauge boson $Z_\mu^{(1)}$ to fermions become
larger.  Anomaly coefficients associated with $Z_\mu^{(1)}$ become larger
as the warp factor $z_L$ becomes larger.
Each coefficient has nontrivial $\theta_H$-dependence.

\begin{figure}[bth]
\centering
\includegraphics[height=47mm]{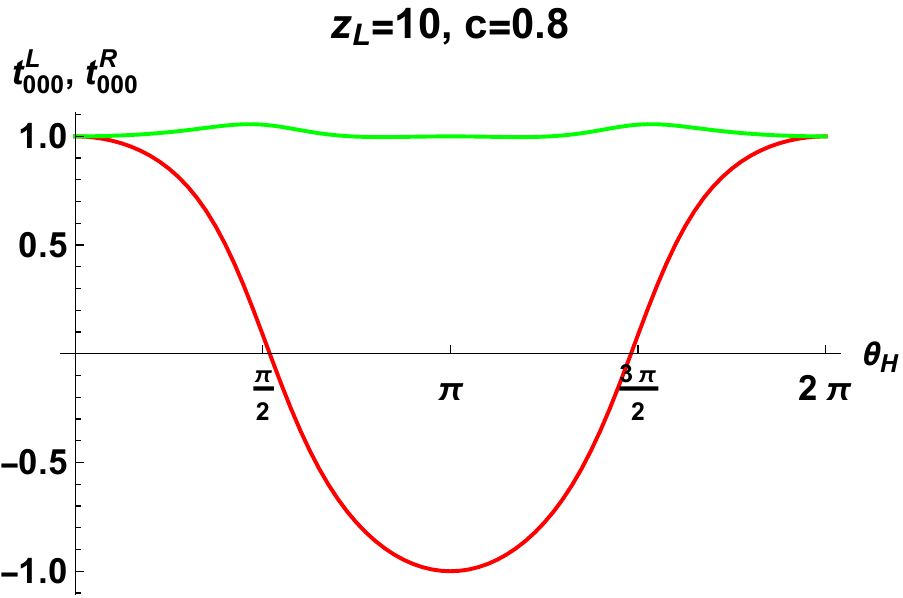}
~
\includegraphics[height=47mm]{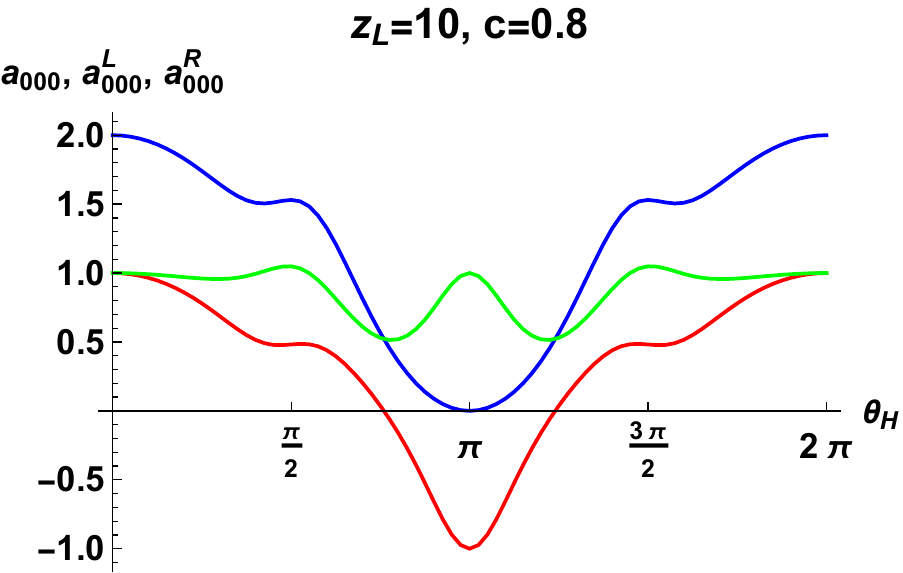}
\caption{Left: The couplings $t_{000}^L$ (green) and $t_{000}^R$ (red).
Right: The anomaly coefficients $a_{000}$ (blue), $a_{000}^L$ (green) and $a_{000}^R$ (red).
Both plots are for $z_L=10$ and $c=0.8$.
$a_{000} (\theta_H)$ shows little dependence on $c$.
}   
\label{fig:anomaly3}
\end{figure}

\begin{figure}[bth]
\centering
\includegraphics[height=45mm]{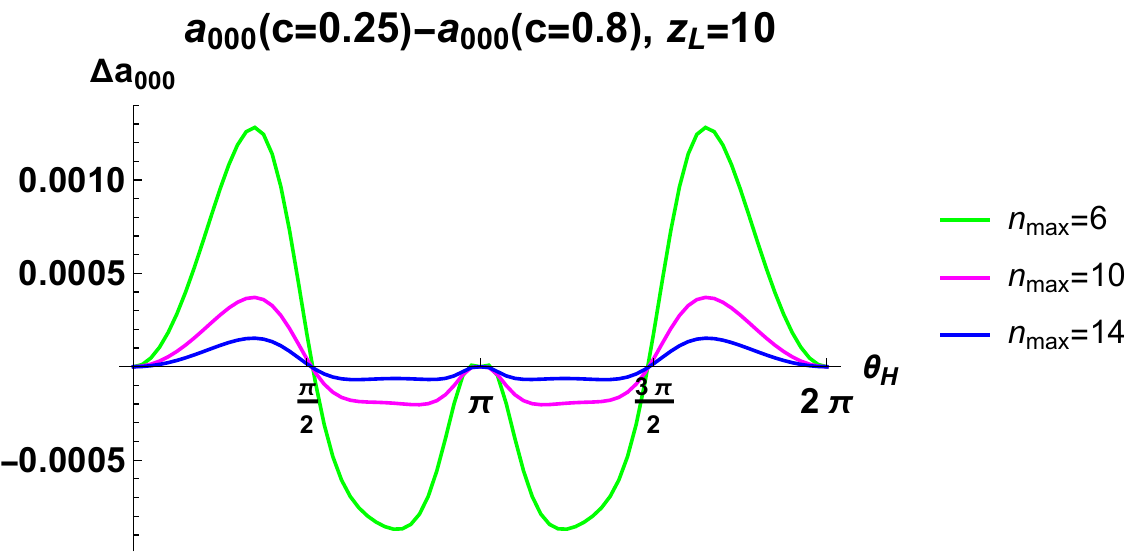}
\caption{The dependence of the anomaly coefficient $a_{000} (\theta_H)$ on the bulk mass parameter $c$.
$a_{000} (\theta_H)_{c=0.25} - a_{000} (\theta_H)_{c=0.8}$ for $z_L=10$ evaluated  by
taking account of fermion modes $\chi^{(n)}$ ($n=0, \cdots, n_\max$) is shown for
$n_\max = 6$ (green), 10 (magenta) and 14 (blue).  The result indicates  $a_{000} (\theta_H)$ becomes 
$c$-independent as $n_\max \go \infty$.
}   
\label{fig:cdependence}
\end{figure}

The anomaly coefficient $a_{n_{1} n_{2} n_{3}} (\theta_{H})$ depends on the warp factor $z_{L}$ and 
the bulk mass parameter $c$ as well.  The couplings $t_{000}^R$ and $t_{000}^L$
and  the anomaly coefficients $a_{000}$, $a_{000}^R$ and  $a_{000}^L$ for $c=0.8$ and $z_L=10$ are
displayed in Fig.\ \ref{fig:anomaly3}.
The couplings of right-handed and left-handed fermions exhibit large $c$-dependence.  
It is seen, however,  that the  total anomaly coefficient $a_{000} (\theta_H)$ is independent of $c$,
being universal.
In the numerical evaluation of anomalies we have incorporated  contributions of fermions 
$\chi^{(n)}$ ($n=0, \cdots, n_\max$).
In Fig.\ \ref{fig:cdependence}, 
$\Delta a_{000} (\theta_H)= a_{000} (\theta_H)_{c=0.25} - a_{000} (\theta_H)_{c=0.8}$ 
is plotted with $n_\max = 6, 10$ and 14 for $z_L=10$.   
As $n_\max$ is increased, the difference $\Delta a_{000} (\theta_H)$
diminishes, approaching to zero.  The maximum of $|\Delta a_{000} (\theta_H)|$ is about 0.000153
at $\theta_H= \frac{7}{20} \pi$ and $\frac{33}{20} \pi$ for $n_\max = 14$.
It is expected that $a_{000} (\theta_H)$ becomes $c$-independent in the $n_\max \go \infty$ limit.
We stress that the $c$-independence of $a_{n \ell m} (\theta_H)$ is highly nontrivial as
the gauge couplings $t_{n \ell m}^{R/L} $ depend on $c$.

For negative $c$ the role of left-handed and right-handed fermions are interchanged.
In other words $t_{n \ell m}^R |_{-c} = t_{n \ell m}^L |_{c}$ and $a_{n \ell m}^R |_{-c} = a_{n \ell m}^L |_{c}$, 
and therefore $a_{n \ell m} (\theta_H)_{-c} = a_{n \ell m}(\theta_H)_{c}$.

\begin{figure}[th]
\centering
\includegraphics[height=47mm]{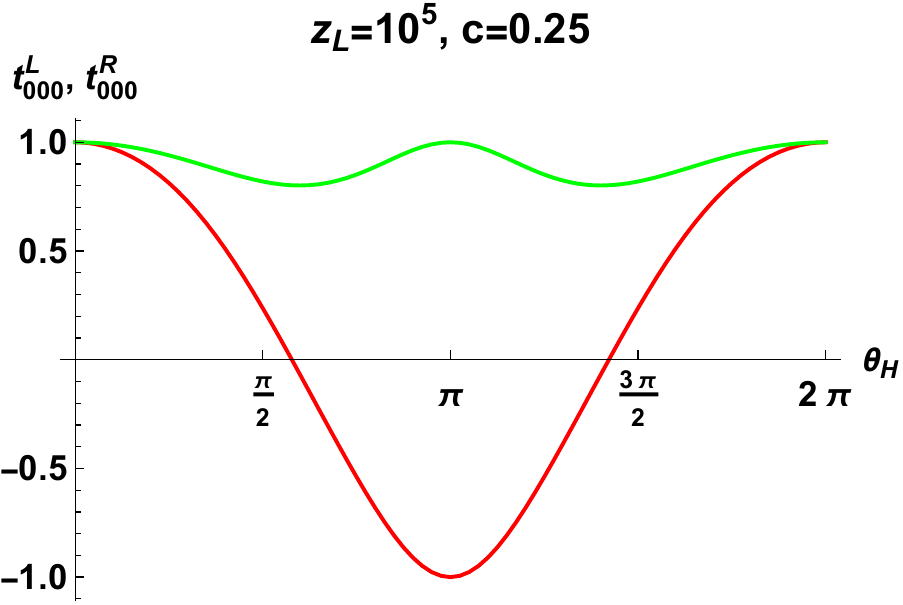}
~
\includegraphics[height=47mm]{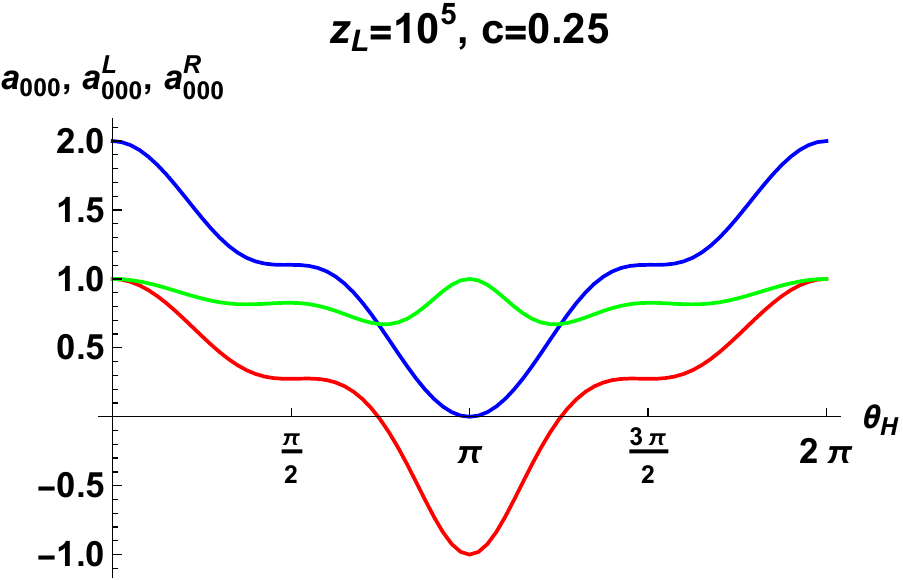}
\caption{Left: The couplings $t_{000}^L$ (green) and $t_{000}^R$ (red).
Right: The anomaly coefficients $a_{000}$ (blue), $a_{000}^L$ (green) and $a_{000}^R$ (red).
Both plots are for $z_L=10^5$ and $c=0.25$.
}   
\label{fig:anomaly4}
\end{figure}

The $z_L$-dependence is investigated similarly.  For large $z_L = e^{kL} \gg 1$ the qualitative behavior does not
change very much.  In Fig.\ \ref{fig:anomaly4} the couplings $t_{000}^R$ and $t_{000}^L$
and  the anomaly coefficients $a_{000}$, $a_{000}^R$ and  $a_{000}^L$ for $z_L=10^{5}$ and $c=0.25$  are
displayed. Compared to the case of $z_L=10$ and $c=0.25$, the behavior of the anomaly coefficients 
becomes milder.

\begin{figure}[bth]
\centering
\includegraphics[height=50mm]{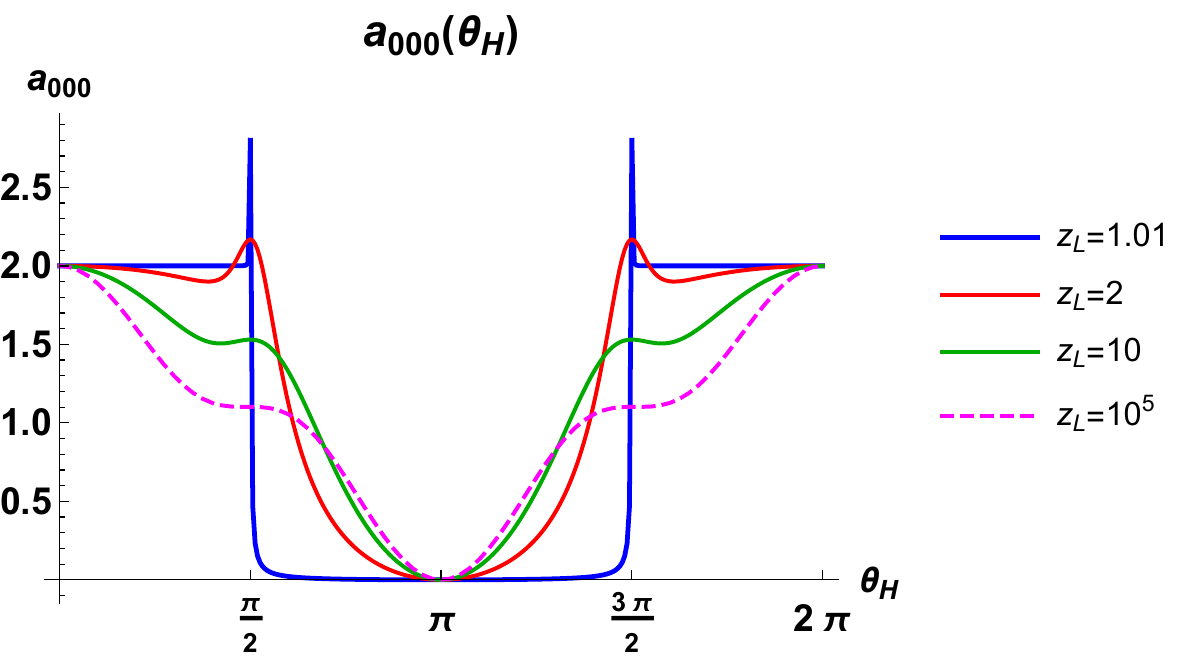}
\vskip 5pt
\includegraphics[height=47mm]{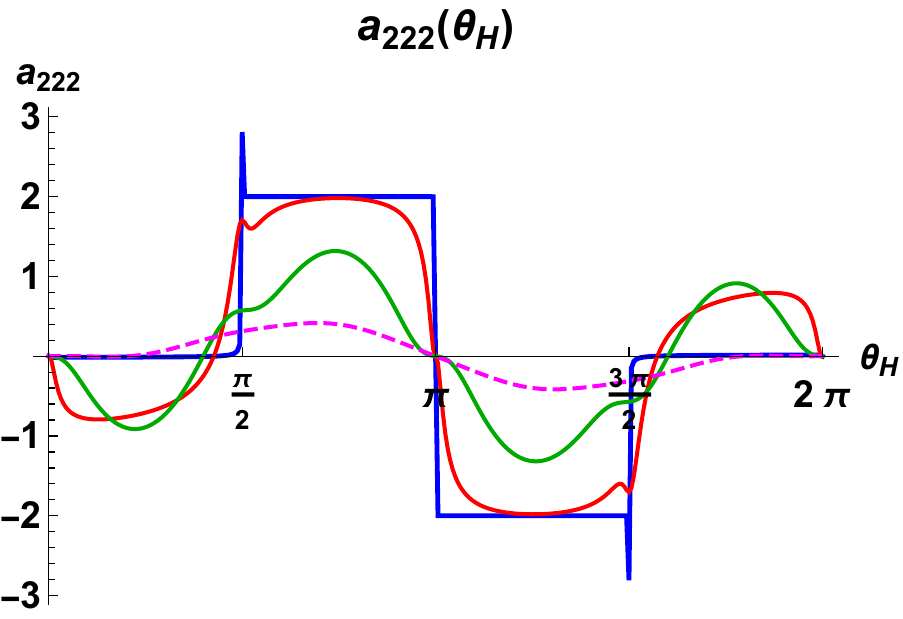} ~
\includegraphics[height=47mm]{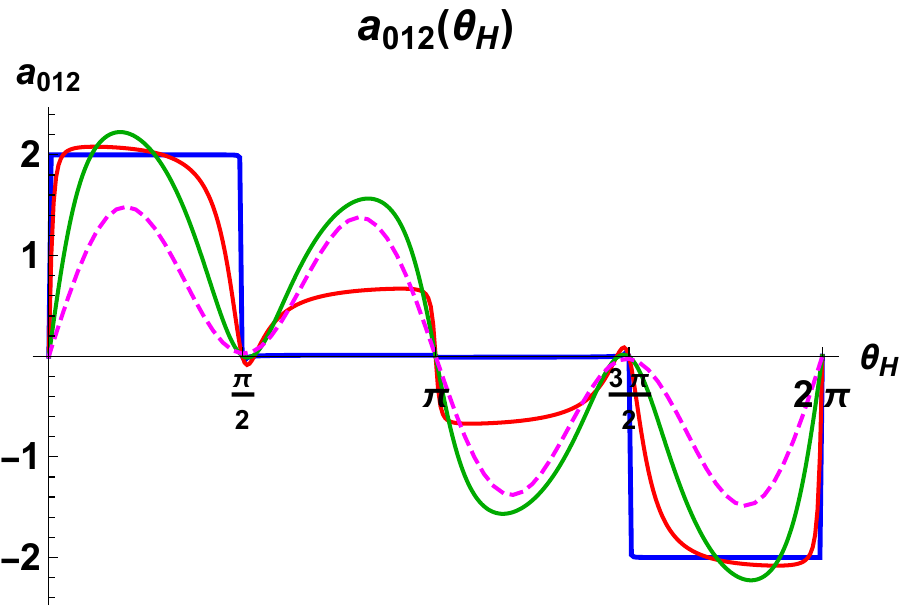} 
\caption{The anomaly coefficients $a_{000}(\theta_H)$, $a_{222}(\theta_H)$ and $a_{012}(\theta_H)$ 
with $c=0.25$ are displayed for  $z_L=1.01, 2, 10$ and $10^5$.
The flat space limit corresponds to $k \go 0$ and $z_L \go 1$.
}   
\label{fig:anomaly-zLdep}
\end{figure}

The flat space limit, $k/m_{\KK} = (z_{L}-1)/\pi \go 0$, exhibits singular behavior.
In the flat space, $M^{4} \times (S^{1}/Z_{2})$, anomaly coefficients $b_{n \ell m}(\theta_{H})$ are constant.
It implies that except at the points of level crossings, $\theta_{H}=0, \pm \onehalf \pi, \pm \pi$, 
$a_{n \ell m}$ must approach to a constant value in the flat space limit, and therefore must show 
step-function type behavior.   In Fig.\ \ref{fig:anomaly-zLdep} the anomaly coefficients $a_{000}(\theta_H)$,  
$a_{222}(\theta_H)$ and $a_{012}(\theta_H)$ with $c=0.25$ are  plotted for various values of $z_L$.  It is seen that 
all of them approach to step functions with singularities at $\theta_H =0,  \onehalf \pi, \pi$ or $\frac{3}{2} \pi$
as $z_{L} \go 1$.

At this juncture it is appropriate to look at the correspondence between $B_{\mu}^{\la n \ra}$
and $Z_{\mu}^{(n)}$  in the flat space limit,
which can be found from the mass spectra displayed in Fig.\ \ref{fig:spectrum1} and Fig.\ \ref{fig:spectrum2}, 
the mode functions  of $B_{\mu}^{\la n \ra}$ in (\ref{gaugeKKflat2}), and the mode functions of
$Z_{\mu}^{(n)}$ in (\ref{RSgaugeKK2}).
The result for $Z_{\mu}^{(n)}$ ($n=0,1,2,3$) is tabulated in Table \ref{Tab:correspondence1}.
With the use of the relationships in Table \ref{Tab:correspondence1}, the anomaly coefficients
in the flat space limit are easily found.  For instance,
\begin{align}
\lim_{k \go 0}  a_{000} (\theta_{H}) &= \begin{cases} 
b_{000} \cr
\frac{1}{2\sqrt{2}} (b_{000} + b_{-1-1-1} + 3 b_{0-1-1} + 3 b_{00-1}) \cr
b_{-1-1-1}  \cr
\frac{1}{2\sqrt{2}} (b_{-1-1-1} + b_{-2-2-2} + 3 b_{-1-2-2} + 3 b_{-1-1-2})  \cr
b_{-2-2-2} 
\end{cases} \cr
\noalign{\kern 5pt}
&=\begin{cases} 
2 &{\rm for~} 0 \le \theta_{H} < \frac{1}{2} \pi \cr
2\sqrt{2}  &{\rm for~}  \theta_{H} = \frac{1}{2} \pi \cr
0  &{\rm for~}  \frac{1}{2} \pi  < \theta_{H} < \frac{3}{2} \pi \cr
2\sqrt{2}  &{\rm for~}  \theta_{H} = \frac{3}{2} \pi \cr
2 &{\rm for~}  \frac{3}{2} \pi < \theta_{H} \le 2 \pi 
\end{cases} ~.
\label{RSAnomaly2}
\end{align}
In Table \ref{Tab:anomaly}, some of the anomaly coefficients in the flat space limit are tabulated.

\begin{table}[htb]
\renewcommand{\arraystretch}{1.4}
\begin{center}
\begin{tabular}{|c||c|c|c|c|}
\hline
$\theta_{H}$  & $Z^{(0)}$ & $Z^{(1)}$ & $Z^{(2)}$ & $Z^{(3)}$\\
\hline \hline
$0$ & $B^{\la 0 \ra}$ & $\frac{1}{\sqrt{2}} (B^{\la 1 \ra} + B^{\la -1 \ra})$ 
& $\frac{1}{\sqrt{2}} (B^{\la 1 \ra} - B^{\la -1 \ra})$ &$\frac{1}{\sqrt{2}} (B^{\la 2 \ra} + B^{\la -2 \ra})$ \\
\hline
$(0 , \frac{1}{2} \pi)$ &$B^{\la 0 \ra}$ & $B^{\la -1 \ra}$ &$B^{\la 1 \ra}$ &$B^{\la -2 \ra}$ \\
\hline
$\frac{1}{2} \pi$ &$\frac{1}{\sqrt{2}} (B^{\la -1 \ra} + B^{\la 0 \ra})$ &$\frac{1}{\sqrt{2}} (B^{\la -1 \ra} - B^{\la 0 \ra})$
&$\frac{1}{\sqrt{2}} (B^{\la -2 \ra} + B^{\la 1 \ra})$ &$\frac{1}{\sqrt{2}} (B^{\la -2 \ra} - B^{\la 1 \ra})$\\
\hline
$(\frac{1}{2} \pi ,  \pi)$ &$B^{\la -1 \ra}$ &$- B^{\la 0 \ra}$ &$B^{\la -2 \ra}$ &$- B^{\la 1 \ra}$\\
\hline
$\pi$ &$B^{\la -1 \ra}$  &$\frac{-1}{\sqrt{2}} (B^{\la -2 \ra} + B^{\la 0 \ra})$
&$\frac{1}{\sqrt{2}} (B^{\la -2 \ra} - B^{\la 0 \ra})$ &$\frac{-1}{\sqrt{2}} (B^{\la -3 \ra} + B^{\la 1 \ra})$\\
\hline
$(\pi ,  \frac{3}{2}  \pi )$ &$B^{\la -1 \ra}$ &$- B^{\la -2 \ra}$ &$-B^{\la 0 \ra}$ &$- B^{\la -3 \ra}$\\
\hline
$\frac{3}{2} \pi$ &$\frac{1}{\sqrt{2}} (B^{\la -1 \ra} + B^{\la -2 \ra})$ &$\frac{1}{\sqrt{2}} (B^{\la -1 \ra} - B^{\la -2 \ra})$
&$\frac{-1}{\sqrt{2}} (B^{\la0 \ra} + B^{\la -3 \ra})$ &$\frac{1}{\sqrt{2}} (B^{\la 0 \ra} - B^{\la -3 \ra})$\\
\hline
$(\frac{3}{2} \pi ,  2 \pi )$ &$B^{\la -2 \ra}$ &$B^{\la -1 \ra}$ &$-B^{\la -3 \ra}$ &$ B^{\la 0 \ra}$\\
\hline
$ 2\pi$ &$B^{\la -2 \ra}$  &$\frac{1}{\sqrt{2}} (B^{\la -1 \ra} + B^{\la -3 \ra})$
&$\frac{1}{\sqrt{2}} (B^{\la -1 \ra} - B^{\la -3 \ra})$ &$\frac{1}{\sqrt{2}} (B^{\la 0 \ra} + B^{\la -4 \ra})$\\
\hline
$(2 \pi ,\frac{5}{2} \pi )$ &$B^{\la -2 \ra}$ & $B^{\la -3 \ra}$ &$B^{\la -1 \ra}$ &$B^{\la -4 \ra}$ \\
\hline
\end{tabular}
\caption{
The correspondence between $B_{\mu}^{\la n \ra}$ and $Z_{\mu}^{(n)}$  in the flat space limit is shown.  
}
\label{Tab:correspondence1}
\end{center}
\end{table}

\begin{table}[hbt]
\renewcommand{\arraystretch}{1.2}
\begin{center}
\begin{tabular}{|c||c|c|c|c|c|c|c|c|}
\hline
$\theta_{H}$  & $a_{000}$ & $a_{111}$ & $a_{222}$ & $a_{001}$ & $a_{011}$ & $a_{002}$  &$a_{022}$ &$a_{012}$\\
\hline \hline
$0$ &2 &0 &0 &0 &4 &0 &0 &0\\
\hline
$(0 , \frac{1}{2} \pi)$ &2 &0 &0 &0 &2 &0 &2 &2\\
\hline
$\frac{1}{2} \pi$ &$2 \sqrt{2}$ &$-2 \sqrt{2}$ &$2 \sqrt{2}$ &0 &0 &$2 \sqrt{2}$ &$2 \sqrt{2}$ &0\\
\hline
$(\frac{1}{2} \pi ,  \pi)$ &0 &$-2$ &2 &$-2$ &0 &2 &0 &0\\
\hline
$\pi$ &0 &$-4 \sqrt{2}$ &0 &$-2 \sqrt{2}$ &0 &0 &0 &0\\
\hline
$(\pi ,  \frac{3}{2}  \pi )$ &0 &$-2$ &$-2$ &$-2$ &0 &$-2$ &0 &0\\
\hline
$\frac{3}{2} \pi$ &$2 \sqrt{2}$ &$-2 \sqrt{2}$ &$- 2 \sqrt{2}$ &0 &0 &$- 2 \sqrt{2}$ &$2 \sqrt{2}$ &0 \\
\hline
$(\frac{3}{2} \pi ,  2 \pi )$ &2 &0 &0 &0 &2 &0 &2 &$-2$ \\
\hline
$ 2\pi$ &2 &0 &0 &0 &4 &0 & 0 &0\\
\hline
\end{tabular}
\caption{
The anomaly coefficients $a_{n \ell m} (\theta_{H})$ due to the $\Psi$ field in the flat space limit are shown.  
Singular behavior is observed at $\theta_{H} = 0, \frac{1}{2} \pi, \pi$ and $\frac{3}{2} \pi$.
}
\label{Tab:anomaly}
\end{center}
\end{table}

The behavior of the anomaly coefficients for $z_{L} = 1.01$  depicted in Fig.\  \ref{fig:anomaly-zLdep}
is understood from the limiting values tabulated in Table \ref{Tab:anomaly}.
In the RS space the anomaly coefficients $a_{n \ell m} (\theta_{H})$ smoothly vary in $\theta_{H}$.
In the flat space limit, however, they exhibit singular behavior at $\theta_{H} = 0, \frac{1}{2} \pi, \pi$ and $\frac{3}{2} \pi$.
This phenomenon is tightly connected with the emergence of the level crossings in the mass spectrum
of the gauge and fermion fields at those points.

Contributions of the $\Psi'$ field to anomalies are evaluated in a similar manner.
The spectrum of the $\Psi'$ field is given by (\ref{RSfermispectrum2}).  
Mode functions ${\bf f}_{Rn}' (z)$ and ${\bf f}_{Ln}' (z)$ are obtained from 
${\bf f}_{Rn} (z)$ and ${\bf f}_{Ln} (z)$ in (\ref{RSfermionKK2R}) and (\ref{RSfermionKK2L}) 
by making a shift $\theta_H \go \theta_H + \pi$.
For instance,  ${\bf f}_{R, 2\ell}' (z)$ is given by 
${\bar{\bf f}}_{R, 2\ell}^{\prime a} (z) = {\bar{\bf f}}_{R, 2\ell}^{b} (z)|_{\theta_H \go \theta_H + \pi}$
for $-\pi < \theta_H < \pi$ and 
${\bar{\bf f}}_{R, 2\ell}^{\prime b} (z) = - {\bar{\bf f}}_{R, 2\ell}^{a} (z)|_{\theta_H \go \theta_H + \pi}$
for $\pi < \theta_H < 2\pi$.
As in the case of the anomaly coefficients $a_{nm\ell}$ coming from the $\Psi$ field,
the anomaly coefficients $a_{nm\ell}'$ coming from the $\Psi'$ field exhibit singular behavior
in the flat space limit. 
In Table \ref{Tab:anomaly2}, some of the anomaly coefficients $a_{nm\ell}'$ in the flat space limit are tabulated.
\begin{table}[hbt]
\renewcommand{\arraystretch}{1.2}
\begin{center}
\begin{tabular}{|c||c|c|c|c|c|c|c|c|}
\hline
$\theta_{H}$  & $a_{000}'$ & $a_{111}'$ & $a_{222}'$ & $a_{001}'$ & $a_{011}'$ & $a_{002}'$  &$a_{022}'$ &$a_{012}'$\\
\hline \hline
$0$ &0 &$4 \sqrt{2}$ &0 &$2 \sqrt{2}$ &0 &0 &0 &0\\
\hline
$(0 ,  \frac{1}{2}  \pi )$ &0 &$2$ &$2$ &$2$ &0 &$2$ &0 &0\\
\hline
$\frac{1}{2} \pi$ &$2 \sqrt{2}$ &$2 \sqrt{2}$ &$2 \sqrt{2}$ &0 &0 &$2 \sqrt{2}$ &$2 \sqrt{2}$ &0 \\
\hline
$(\frac{1}{2} \pi ,  \pi )$ &2 &0 &0 &0 &2 &0 &2 &$-2$ \\
\hline
$\pi$ &2 &0 &0 &0 &4 &0 &0 &0\\
\hline
$(\pi , \frac{3}{2} \pi)$ &2 &0 &0 &0 &2 &0 &2 &2\\
\hline
$\frac{3}{2} \pi$ &$2 \sqrt{2}$ &$2 \sqrt{2}$ &$-2 \sqrt{2}$ &0 &0 &$-2 \sqrt{2}$ &$2 \sqrt{2}$ &0\\
\hline
$(\frac{3}{2} \pi ,  2\pi)$ &0 &$2$ &$-2$ &$2$ &0 &$-2$ &0 &0\\
\hline
$2\pi$ &0 &$4 \sqrt{2}$ &0 &$2 \sqrt{2}$ &0 &0 &0 &0\\
\hline
\end{tabular}
\caption{
The anomaly coefficients $a_{n \ell m}' (\theta_{H})$ due to the $\Psi'$ field in the flat space limit are shown.  
Singular behavior is observed at $\theta_{H} = 0, \frac{1}{2} \pi, \pi$ and $\frac{3}{2} \pi$.
$a_{n \ell m}' (\theta_{H})$ is related to $a_{n \ell m} (\theta_{H})$ in Table \ref{Tab:anomaly}
by $a_{n \ell m}' (\theta_{H}) =  a_{n \ell m} (\theta_{H} + \pi)$ or $- a_{n \ell m} (\theta_{H} + \pi)$.
}
\label{Tab:anomaly2}
\end{center}
\end{table}

\section{Summary and discussion} 

In this paper we have shown that chiral triangle anomalies smoothly flow 
in the scheme of $SU(2)$ GHU  models in the RS space 
as the AB phase $\theta_H$ in the fifth dimension varies.
Zero modes of $SU(2)$ doublet fermions $\Psi$  have  chiral gauge couplings at $\theta_H=0$.
Those gauge couplings smoothly change as $\theta_H$, and they become vectorlike at $\theta_H = \pi$.
Although everything changes smoothly in the RS space, the flat space limit becomes singular
at $\theta_H = 0, \frac{1}{2} \pi, \pi$ and $\frac{3}{2} \pi$ where the level crossings in the mass
spectrum occur in the flat $M^4 \times (S^1/Z_2)$ spacetime.

The anomaly coefficients $a^R_{n \ell m} (\theta_{H})$ and $a^L_{n \ell m} (\theta_{H})$ in the RS space 
depend on the warp factor $z_L$ and  the bulk mass parameter $c$ of the fermion field.  
We have confirmed by numerical evaluation that the total anomaly coefficients 
$a_{n \ell m} (\theta_{H}) = a^R_{n \ell m} (\theta_{H}) + a^L_{n \ell m} (\theta_{H}) $ are
independent of the value of $c$. 
This may have  profound implications for realistic GHU models in the RS space.
In the $SO(5) \times U(1) \times SU(3)$ GHU \cite{HNU2010, Amodel2013, GUTinspired2019},
for instance, quark-lepton multiplets are introduced
such that all gauge anomalies are cancelled at $\theta_H=0$. 
Each quark or lepton multiplet has its own bulk mass parameter $c$. 
In the vacuum  $\theta_H \not= 0$ and the electroweak symmetry is dynamically broken. 
Typically $\theta_H \sim  0.1$ and $z_L = 10^5 \sim 10^{10}$.  
Gauge couplings of right- and left-handed quark or lepton
change slightly at $\theta_H \not= 0$, depending on $c$.  The universality of $a_{n \ell m} (\theta_{H})$
implies that all gauge anomalies remain cancelled even at $\theta_H \not= 0$.\cite{Bouchiat1972, GrossJackiw1972}

Anomalies flow by an AB phase.  It is known that anomalies in four dimensions are related to
global topology of the space through the index theorem.\cite{AtiyahSinger1968, AtiyahPatodiSinger1975}
It is challenging to understand the anomaly flow
by an AB phase from the viewpoint of  the index theorem.\cite{Fukaya2017, WittenYonekura2021}
Gauge theory in the RS space or in the flat $M^4 \times (S^1/Z_2)$ spacetime can be formulated
as gauge theory on an interval $(0 \le y \le L)$ in the fifth dimension
with a special class of orbifold boundary conditions at $y=0$ and $L$.
In the twisted gauge in GHU the AB phase $\theta_H$ appears as a phase parameter specifying orbifold 
boundary conditions.  Anomaly and the index theorem in orbifold gauge theory with nonvanishing $\theta_H$ 
have not been well understood so far.
To elucidate the anomaly flow by $\theta_H$  the RS space will provide a powerful tool 
as there occurs no level crossing in the mass spectrum and anomaly smoothly changes as $\theta_H$, 
quite in contrast to the behavior in the flat space.

\section*{Acknowledgements}

This work was supported in part by European Regional Development Fund-Project Engineering Applications 
of Microworld Physics (Grant No. CZ.02.1.01/0.0/0.0/16$\underline{~}$019/0000766) (Y.O.), 
and by Japan Society for the Promotion of Science, Grants-in-Aid for Scientific 
Research, Grant No. JP19K03873 (Y.H.) and Grant No. JP18H05543  (N.Y.).

\vskip 1.cm

\appendix
\section{Basis functions} 
Wave functions of gauge fields and fermions are expressed in terms of the following basis functions.
For gauge fields we introduce
\begin{align}
 C(z; \lambda) &= \frac{\pi}{2} \lambda z z_L F_{1,0}(\lambda z, \lambda z_L) ~,  \cr
 S(z; \lambda) &= -\frac{\pi}{2} \lambda  z F_{1,1}(\lambda z, \lambda z_L) ~, \cr
 C^\prime (z; \lambda) &= \frac{\pi}{2} \lambda^2 z z_L F_{0,0}(\lambda z, \lambda z_L) ~,  \cr
S^\prime (z; \lambda) &= -\frac{\pi}{2} \lambda^2 z  F_{0,1}(\lambda z, \lambda z_L)~, \cr
\noalign{\kern 5pt}
 F_{\alpha, \beta}(u, v) &\equiv 
J_\alpha(u) Y_\beta(v) - Y_\alpha(u) J_\beta(v) ~,
\label{functionA1}
\end{align}
where $J_\alpha (u)$ and $Y_\alpha (u)$ are Bessel functions of  the first and second kind.
They satisfy
\begin{align}
&- z \frac{d}{dz} \frac{1}{z} \frac{d}{dz} \begin{pmatrix} C \cr S \end{pmatrix} 
= \lambda^{2} \begin{pmatrix} C \cr S \end{pmatrix} ~, \cr
\noalign{\kern 5pt}
&C(z_{L} ; \lambda)  = z_{L} ~, ~~ C' (z_{L} ; \lambda)  = 0 ~, \cr
\noalign{\kern 5pt}
&S(z_{L} ; \lambda)  = 0 ~, ~~ S' (z_{L} ; \lambda)  = \lambda ~,  \cr
\noalign{\kern 5pt}
&CS' - S C' = \lambda z ~.
\label{relationA1}
\end{align}
To express wave functions of KK modes of gauge fields,  we  make use of
\begin{align}
&\hat S(z;\lambda) = N_{0}(\lambda) S(z;\lambda)  ~,~~
\hat C(z;\lambda) = N_{0}(\lambda)^{-1} C(z;\lambda)  ~, \cr
\noalign{\kern 5pt}
&\check S(z;\lambda) = N_{1}(\lambda) S(z;\lambda)  ~,~~
\check C(z;\lambda) = N_{1}(\lambda)^{-1} C(z;\lambda)  ~, \cr
\noalign{\kern 5pt}
&\qquad N_{0}(\lambda) = \frac{C(1;\lambda)}{S(1;\lambda)} ~, ~~
N_{1}(\lambda) = \frac{C'(1;\lambda)}{S'(1;\lambda)} ~.
\label{functionA2}
\end{align}

For fermion fields with a bulk mass parameter $c$, we define 
\begin{align}
\begin{pmatrix} C_L \cr S_L \end{pmatrix} (z; \lambda,c)
&= \pm \frac{\pi}{2} \lambda \sqrt{z z_L} F_{c+\frac12, c\mp\frac12}(\lambda z, \lambda z_L) ~, \cr
\begin{pmatrix} C_R \cr S_R \end{pmatrix} (z; \lambda,c)
&= \mp \frac{\pi}{2} \lambda \sqrt{z z_L} F_{c- \frac12, c\pm\frac12}(\lambda z, \lambda z_L) ~.
\label{functionA3}
\end{align}
These functions satisfy 
\begin{align}
&D_{+} (c) \begin{pmatrix} C_{L} \cr S_{L} \end{pmatrix} = \lambda  \begin{pmatrix} S_{R} \cr C_{R} \end{pmatrix}, \cr
\noalign{\kern 5pt}
&D_{-} (c) \begin{pmatrix} C_{R} \cr S_{R} \end{pmatrix} = \lambda  \begin{pmatrix} S_{L} \cr C_{L} \end{pmatrix}, ~~
D_{\pm} (c) = \pm \frac{d}{dz} + \frac{c}{z} ~, \cr
\noalign{\kern 5pt}
&C_{R} = C_{L} = 1 ~, ~~ S_{R} = S_{L} = 0 \quad {\rm at~} z=z_{L} ~, \cr
\noalign{\kern 5pt}
&C_L C_R - S_L S_R=1 ~. 
\label{relationA2}
\end{align}
Also $C_L  (z; \lambda, -c) = C_R  (z; \lambda, c)$ and $S_L  (z; \lambda, -c) = - S_R  (z; \lambda, c)$ hold.  
To express wave functions of KK modes of fermion fields,  we  make use of
\begin{align}
&\hat S_{L}  (z; \lambda,c) = N_{L}( \lambda,c) S_{L}  (z; \lambda,c) ~, ~~
\hat C_{L}  (z; \lambda,c) = N_{R}( \lambda,c) C_{L}  (z; \lambda,c) ~,  \cr
\noalign{\kern 5pt}
&\hat S_{R}  (z; \lambda,c) = N_{R}( \lambda,c) S_{R}  (z; \lambda,c) ~, ~~
\hat C_{R}  (z; \lambda,c) = N_{L}( \lambda,c) C_{R}  (z; \lambda,c) ~,  \cr
\noalign{\kern 5pt}
&\check S_{L}  (z; \lambda,c) = N_{R}( \lambda,c)^{-1} S_{L}  (z; \lambda,c) ~, ~~
\check C_{L}  (z; \lambda,c) = N_{L}( \lambda,c)^{-1} C_{L}  (z; \lambda,c) ~,  \cr
\noalign{\kern 5pt}
&\check S_{R}  (z; \lambda,c) = N_{L}( \lambda,c)^{-1} S_{R}  (z; \lambda,c) ~, ~~
\check C_{R}  (z; \lambda,c) = N_{R}( \lambda,c)^{-1} C_{R}  (z; \lambda,c) ~,  \cr
\noalign{\kern 5pt}
&\qquad N_{L}( \lambda,c)  = \frac{C_{L} (1; \lambda,c)}{S_{L} (1; \lambda,c)} ~, ~~
N_{R}( \lambda,c)  = \frac{C_{R} (1; \lambda,c)}{S_{R} (1; \lambda,c)} ~.
\label{functionA4}
\end{align}

\section{Mode functions of fermion fields with $c<0$} 

When the bulk mass parameter $c$ of a fermion field $\Psi (x,z)$ is negative, the roles of right-handed
and left-handed components are interchanged compared with those of a field with a positive $c$.
In the KK expansions (\ref{RSfermionKK1})  mode functions ${\bf f}_{Rn} (z)$ and  ${\bf f}_{Ln} (z)$
are given, for $c <0$, by

\begin{align}
{\bf f}_{R0} (z)  &=  \bar {\bf f}_{R0}^a (z) , \cr
\noalign{\kern 5pt}
{\bf f}_{R, 2\ell-1} (z)  &= 
\begin{cases} \bar {\bf f}_{R, 2\ell-1}^a (z) &{\rm for~} - \pi < \theta_H < \pi  \cr
\bar {\bf f}_{R, 2\ell-1}^b (z) &{\rm for~} 0 < \theta_H <  2\pi \cr
- \bar {\bf f}_{R, 2\ell-1}^a (z) &{\rm for~}   \pi < \theta_H < 3\pi  \cr
- \bar {\bf f}_{R, 2\ell-1}^b (z) &{\rm for~} 2 \pi < \theta_H <  4\pi \cr
\bar {\bf f}_{R, 2\ell-1}^a (z) &{\rm for~}  3 \pi < \theta_H < 5 \pi  \cr \end{cases} ~ (\ell=1,2,3, \cdots) , \cr
\noalign{\kern 5pt}
{\bf f}_{R, 2\ell} (z)  &= 
\begin{cases} \bar {\bf f}_{R, 2\ell}^c (z) &{\rm for~} - \pi < \theta_H < \pi  \cr
\bar {\bf f}_{R, 2\ell}^d (z) &{\rm for~} 0 < \theta_H <  2\pi \cr
- \bar {\bf f}_{R, 2\ell}^c (z) &{\rm for~}   \pi < \theta_H < 3\pi  \cr
- \bar {\bf f}_{R, 2\ell}^d (z) &{\rm for~} 2 \pi < \theta_H <  4\pi \cr
\bar {\bf f}_{R, 2\ell}^c (z) &{\rm for~}  3 \pi < \theta_H < 5 \pi  \cr \end{cases} ~ (\ell=1,2, 3, \cdots),
\label{RSfermionKK4R}
\end{align}
and 
\begin{align}
{\bf f}_{L, 2\ell} (z)  &= 
\begin{cases} \bar {\bf f}_{L, 2\ell}^a (z) &{\rm for~} - \pi < \theta_H < \pi  \cr
\bar {\bf f}_{L, 2\ell}^b (z) &{\rm for~} 0 < \theta_H <  2\pi \cr
- \bar {\bf f}_{L, 2\ell}^a (z) &{\rm for~}   \pi < \theta_H < 3\pi  \cr
- \bar {\bf f}_{L, 2\ell}^b (z) &{\rm for~} 2 \pi < \theta_H <  4\pi \cr
\bar {\bf f}_{L, 2\ell}^a (z) &{\rm for~}  3 \pi < \theta_H < 5 \pi  \cr \end{cases} ~ (\ell=0, 1,2,\cdots), \cr
\noalign{\kern 5pt}
{\bf f}_{L, 2\ell-1} (z)  &= 
\begin{cases} \bar {\bf f}_{L, 2\ell-1}^c (z) &{\rm for~} - \pi < \theta_H < \pi  \cr
\bar {\bf f}_{L, 2\ell-1}^d (z) &{\rm for~} 0 < \theta_H <  2\pi \cr
- \bar {\bf f}_{L, 2\ell-1}^c (z) &{\rm for~}   \pi < \theta_H < 3\pi  \cr
- \bar {\bf f}_{L, 2\ell-1}^d (z) &{\rm for~} 2 \pi < \theta_H <  4\pi \cr
\bar {\bf f}_{L, 2\ell-1}^c (z) &{\rm for~}  3 \pi < \theta_H < 5 \pi  \cr \end{cases} ~ (\ell=1,2,3, \cdots) .
\label{RSfermionKK4L}
\end{align}
Here $\bar {\bf f}_{Rn}^a (z)$, $\bar {\bf f}_{Rn}^b (z)$ etc. are given in (\ref{RSfermionKK3}).

\vskip 1.cm

\def\jnl#1#2#3#4{{#1}{\bf #2},  #3 (#4)}

\def\Zphys{{\em Z.\ Phys.} }
\def\jssc{{\em J.\ Solid State Chem.\ }}
\def\jpsJ{{\em J.\ Phys.\ Soc.\ Japan }}
\def\ptps{{\em Prog.\ Theoret.\ Phys.\ Suppl.\ }}
\def\PTP{{\em Prog.\ Theoret.\ Phys.\  }}
\def\PTEP{{\em Prog.\ Theoret.\ Exp.\  Phys.\  }}
\def\JMP{{\em J. Math.\ Phys.} }
\def\NPB{{\em Nucl.\ Phys.} B}
\def\NP{{\em Nucl.\ Phys.} }
\def\PLB{{\it Phys.\ Lett.} B}
\def\PL{{\em Phys.\ Lett.} }
\def\PRL{\em Phys.\ Rev.\ Lett. }
\def\PRB{{\em Phys.\ Rev.} B}
\def\PRD{{\em Phys.\ Rev.} D}
\def\PRe{{\em Phys.\ Rep.} }
\def\AP{{\em Ann.\ Phys.\ (N.Y.)} }
\def\RMP{{\em Rev.\ Mod.\ Phys.} }
\def\ZPC{{\em Z.\ Phys.} C}
\def\SCI{\em Science}
\def\CMP{\em Comm.\ Math.\ Phys. }
\def\MPLA{{\em Mod.\ Phys.\ Lett.} A}
\def\IJMPA{{\em Int.\ J.\ Mod.\ Phys.} A}
\def\IJMPB{{\em Int.\ J.\ Mod.\ Phys.} B}
\def\EPJC{{\em Eur.\ Phys.\ J.} C}
\def\PR{{\em Phys.\ Rev.} }
\def\JHEP{{\em JHEP} }
\def\JCAP{{\em JCAP} }
\def\cmp{{\em Com.\ Math.\ Phys.}}
\def\JPA{{\em J.\  Phys.} A}
\def\JPG{{\em J.\  Phys.} G}
\def\NJP{{\em New.\ J.\  Phys.} }
\def\CQG{\em Class.\ Quant.\ Grav. }
\def\ATMP{{\em Adv.\ Theoret.\ Math.\ Phys.} }
\def\ibid{{\em ibid.} }
\def\ChP{{\em Chin.Phys.}C}
\def\NCA{{\it Nuovo Cim.} A}


\renewenvironment{thebibliography}[1]
         {\begin{list}{[$\,$\arabic{enumi}$\,$]}  
         {\usecounter{enumi}\setlength{\parsep}{0pt}
          \setlength{\itemsep}{0pt}  \renewcommand{\baselinestretch}{1.2}
          \settowidth
         {\labelwidth}{#1 ~ ~}\sloppy}}{\end{list}}

\leftline{\Large \bf References}


\end{document}